\documentclass[aps,prl,twocolumn,preprintnumbers,amsmath,amssymb,superscriptaddress,bibnotes]{revtex4}
\usepackage{graphicx}
\usepackage{dcolumn}
\usepackage{bm}
\usepackage{color}
\begin{document}

\title{Multiple quantum phase transitions and superconductivity in Ce-based heavy fermions}

\author{Z. F. Weng}
\author{M. Smidman}
\affiliation {Center for Correlated Matter and Department of Physics, Zhejiang University, Hangzhou 310058, China}
\author{L. Jiao}
\affiliation {Center for Correlated Matter and Department of Physics, Zhejiang University, Hangzhou 310058, China}
\author{Xin Lu}
\email{xinluphy@zju.edu.cn}
\author{H. Q. Yuan}
\email{hqyuan@zju.edu.cn}
\affiliation {Center for Correlated Matter and Department of Physics, Zhejiang University, Hangzhou 310058, China}
\affiliation{Collaborative Innovation Center of Advanced Microstructures, Nanjing 210093, China}

\date{\today}

\begin{abstract}
Heavy fermions have served as prototype examples of strongly-correlated electron systems. The occurrence of unconventional superconductivity in close proximity to the electronic instabilities associated with various degrees of freedom points to an intricate relationship between superconductivity and other electronic states, which is unique but also shares some common features with high temperature superconductivity. The magnetic order in heavy fermion compounds can be continuously suppressed by tuning external parameters to a quantum critical point, and the role of quantum criticality in determining the properties of heavy fermion systems is an important unresolved issue. Here we review the recent progress of studies on Ce based heavy fermion superconductors, with an emphasis on the superconductivity emerging on the edge of magnetic and charge instabilities as well as the quantum phase transitions which occur by tuning different parameters, such as pressure, magnetic field and doping. We discuss systems where multiple quantum critical points occur and whether they can be classified in a unified manner, in particular in terms of the evolution of the Fermi surface topology.

\end{abstract}

\maketitle
\section{Introduction}
As prototype examples of strongly correlated electron systems, heavy fermions have been intensively investigated since their discovery due to their exotic properties and fascinating underlying physics \cite{GStewartHFReview84,StewartNFL,RevModPhys.79.1015,HFSC09RMP,White2015PhysicaC,Si1161,HewsonHF,PColeman07Review}.  The discovery of unconventional superconductivity in the heavy fermion compound CeCu$_2$Si$_2$ with $T_c\simeq 0.6$~K by Frank Steglich in 1979 \cite{FSteglich79PRL} pioneered research into unconventional superconductivity and in particular, the intricate relationship between superconductivity and magnetism, which had been previously considered to be antagonistic \cite{White2015PhysicaC, HFSC09RMP}. Understanding the close relationship between superconductivity and quantum criticality may be important for solving the puzzles of high-temperature superconductivity in cuprate or iron pnictide superconductors, but its underlying mechanism still remains as an intriguing open issue \cite{MathurQCP98Nature, DPines07Nature, Norman11Science, DScalapino12RMP}.

\begin{figure}[t]
\includegraphics[angle=0,width=0.48\textwidth]{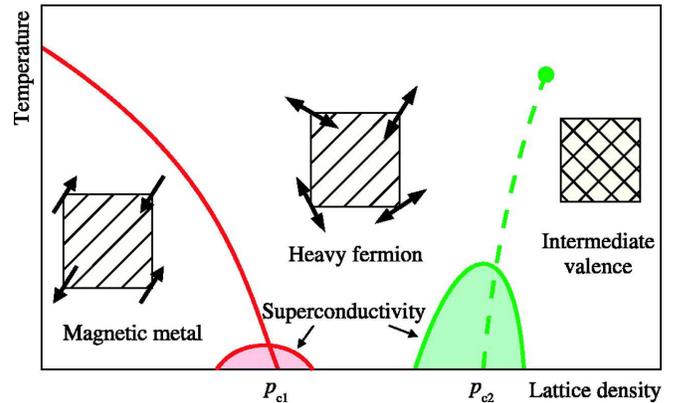}

\caption{ Schematic phase diagram of Ce-based narrow band metals. Upon increasing the hybridization by applying pressure, the ground state changes from an antiferromagnetic to heavy fermion state, before entering the intermediate valence regime. Two critical points associated with spin and valence instabilities occur at $p_{c1}$ and $p_{c2}$ respectively, around which there is unconventional superconductivity. Reprinted with permission from Ref.~\cite{Yuan19122003}. Copyright 2003 AAAS. }\label{F1yuan}
\end{figure}

Heavy fermion systems generally refer to certain intermetallic compounds containing either lanthanide (Ce, Yb, Pr, etc.) or actinide (U, Pu, Np, etc.) elements. Electrons from partially filled $4f$ or $5f$ orbits are arranged in a periodic lattice and interact with the itinerant conduction electrons, which leads to  many emergent quantum phenomena. Among them, Ce-based compounds serve as the simplest examples to demonstrate much of the important heavy fermion physics, since each Ce atom has only one $f$ electron, with an atomic configuration $4f^1(5d6s^2)$. Elemental Ce shows a discontinuous valence transition from a Ce valence of 3.03 ($\gamma$ phase) to 3.14 ($\alpha$ phase) at 300~K \cite{CeValenceTransition84}, while Ce-based materials display a variety of ground states such as magnetic ordering, heavy Fermi liquid state or mixed-valence state. Due to the relatively small energy scales in heavy fermion systems, the ground state  can often be readily tuned by external parameters such as  pressure, magnetic fields or chemical doping. For example, the antiferromagnetism (AFM) can be continuously suppressed  by applying pressure or a magnetic field in CeRhIn$_5$ \cite{park_hidden_2006, PhysRevB.74.020501,Jiao20012015}, pressure in CePd$_2$Si$_2$ \cite{MathurQCP98Nature} or by doping in CeCu$_{6-x}$Au$_x$ \cite{stockert_unconventional_2011}. Moreover, charge degrees of freedom increasingly need to be taken into account upon enhancing the hybridization between the $f$ and conduction electrons, leading to valence instabilities or crossover behavior, as shown in the schematic phase diagram in Fig.~\ref{F1yuan} \cite{Yuan19122003}. Unconventional superconductivity (SC) occurs when the system is on the verge of either magnetic or valence-instabilities \cite{Yuan19122003}, suggesting a close relationship between the SC and the associated critical fluctuations. Presently, almost half of the heavy fermion superconductors which have been found are Ce-based compounds, some of which are superconducting at ambient pressure, while others require high pressures to be applied.

Studies of quantum criticality and superconductivity raise a series of important questions with regards to strongly correlated electron systems: (1) whether there exist distinctly different types of  quantum critical points (QCP) and whether these can be universally classified; (2) the possibility of multiple QCPs upon tuning with different parameters and the validity of the global phase diagram; (3) the underlying pairing mechanisms of heavy fermion superconductivity and its relationship to the QCP. The remainder of the review is  organised as follows. Firstly we will give a brief overview of various aspects of heavy fermion physics and quantum criticality. We will then discuss  examples of systems which become superconducting near an antiferromagnetic instability before discussing the possibility of superconductivity occurring in proximity to valence transitions. Here we present some examples of systems where a parameter such as pressure may tune the system through multiple instabilities, corresponding to different degrees of freedom. We then review some materials which display multiple quantum phase transitions, where distinctly different behaviors are seen when the system is tuned by different parameters. Finally we will discuss to what extent these behaviors can be classified in a universal manner and the applicability of the recently proposed global phase diagram \cite{Si2006Global,SiPSSGlobal,Coleman2010Global} for heavy fermion systems.

\section{Theoretical approaches from heavy fermion to mixed valence state}

\subsection{Kondo vs. RKKY Interaction}
Many Ce based systems can be modelled as a lattice of $f$ electrons, which antiferromagnetically  couple with the conduction electrons via the Kondo interaction. This is the Kondo lattice model given by \cite{DONIACH1977}

\begin{equation}
H=\sum_{k\sigma}{\epsilon_kc_{k\sigma}^{+}c_{k\sigma}}+J\sum_{i}{\overrightarrow{\sigma}(i)\cdot \overrightarrow{S}_{i}};
\label{KLEq}
\end{equation}

\noindent where  $J$ characterizes the coupling strength between the spin density of the conduction electrons $\overrightarrow{\sigma}(i)$ and the $f$~electron spin $\overrightarrow{S}_{i}$ at the site $i$. In the canonical Kondo impurity problem, as the temperature reduces the localized $f$ moments are compensated by the opposite spins of the conduction electrons. Below the characteristic Kondo temperature $T_K$, a Kondo singlet state forms \cite{Kondo1964, HewsonHF}, giving rise to a logarithmic increase in the resistivity due to the enhanced scattering of conduction electrons. The effect of the $f$ electrons being arranged in a periodic lattice, means that upon reducing the temperature further, the electrical resistivity $\rho(T)$ typically reaches a maximum, before decreasing rapidly. This characteristic temperature is commonly called the coherence temperature, below which Kondo scattering becomes coherent and the system  evolves into a state that shows Landau Fermi liquid behavior \cite{LACROIX1986145,PColeman07Review}. However, the renormalized electron mass is much larger than that of free electrons due to the hybridization between localized electrons and conduction electrons. The term ``heavy fermion" has therefore been coined to distinguish the behavior of these materials from that of a normal Fermi liquid in metals \cite{GStewartHFReview84}.

Besides the Kondo interaction, there also exists a competing interaction in heavy fermion systems, the Ruderman-Kittel-Kasuya-Yosida (RKKY) interaction \cite{RKKY1,RKKY2,RKKY3}. While the Kondo interaction screens the magnetic moments of localized $f$ electrons and leads to the formation of a heavy Fermi-liquid state, the RKKY interaction favors long-range magnetic order. Here the localized electrons can polarize the surrounding conduction electrons to form a spin density wave in space, which interacts with other $f$ electrons leading to an ordered state. The temperatures characterizing the Kondo interaction ($T_K$) and  RKKY interaction ($T_{RKKY}$) can both be expressed in terms of the interaction $J$ between the $f$ and conduction electrons and the density of states at the Fermi energy $D(E_{F})$, but they follow different behaviors \cite{DONIACH1977}. The Kondo temperature scales as
\begin{equation}
T_{K} \propto \exp(-\frac{1}{JD(E_{F})});
\end{equation}
while the RKKY temperature follows
\begin{equation}
T_{RKKY} \propto D(E_{F})J^{2}.
\end{equation}

\subsection{Magnetic quantum criticality}

\begin{figure}[t]
\includegraphics[angle=0,width=0.49\textwidth]{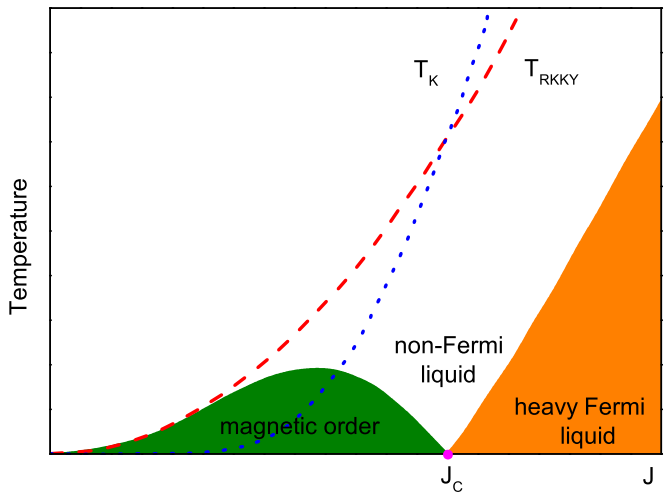}

\caption{\label{FigDoniachPD} A schematic plot of the Doniach phase diagram for heavy fermion systems \cite{DONIACH1977}, where there is competition between the RKKY interaction and Kondo interaction, with relative strengths $T_{RKKY}$ and $T_K$ respectively.}
\end{figure}

The competition between the Kondo and RKKY interactions in heavy fermion systems may result in various ground states, leading to a rich phase diagram, which can be qualitatively described by the Doniach phase diagram as schematically shown in Fig.~\ref{FigDoniachPD} \cite{DONIACH1977}. The hybridization constant $J$ can be effectively tuned by pressure, chemical doping or magnetic field because the associated energy scales are relatively small in heavy fermion compounds. When $J$ is small, the RKKY interaction becomes dominant and the system exhibits a magnetically ordered ground state, such as antiferromagnetism. With increasing hybridization, the Kondo interaction is enhanced and the system may exhibit heavy fermion behavior. At a critical value $J_c$, the system may experience a continuous phase transition at zero temperature, $\textit{i.e.}$, a QCP, changing the ground state from a magnetically ordered state to a disordered state, which is driven by quantum fluctuations instead of thermal fluctuations. Near the quantum critical point of heavy fermion compounds, exotic quantum states and anomalous normal-state behavior have been widely observed \cite{GStewartHFReview84,RevModPhys.79.1015,StewartNFL,HFSC09RMP, White2015PhysicaC, Si1161, HewsonHF, PColeman07Review}. For example, the specific heat $C/T$ can diverge logarithmically and the resistivity may show a linear rather than quadratic temperature dependence as the temperature approaches zero.

The universality of the critical behaviors and its underlying mechanism remain unresolved. It was proposed that QCPs in heavy fermion compounds  can be categorized into two types: conventional spin-density-wave \cite{PhysRevB.14.1165Hertz, PhysRevB.48.7183Millis, Moriya}  or unconventional Kondo breakdown (local) QCPs \cite{si_locally_2001, ColemanPepin, Pepin2007}, where in the latter case, the Fermi surface topology should be taken into consideration in addition to order parameter fluctuations. In the Kondo lattice, whether $f$ electrons contribute to the Fermi surface volume depends on the degree of Kondo screening. There exists a characteristic energy scale line E$_{loc}^*$ at low temperatures, signaling a change from a small Fermi surface volume to a large Fermi surface volume, which is a crossover at finite temperatures, but a sharp transition at zero temperature. In the case of a Kondo-breakdown QCP, the E$_{loc}^*$ line terminates exactly where the ordering temperature $T_{N}$ vanishes and as a result, the critical fluctuations are associated with not only the order parameter but also drastic changes of the  Fermi surface volume \cite{si_locally_2001, ColemanPepin}. On the other hand, when  E$_{loc}^*$ and $T_{N}$ intersect at a finite temperature, the long-range magnetic order switches from local moment magnetism, with a small Fermi surface volume(AF$_S$) to an itinerant spin-density-wave with a large Fermi surface volume (AF$_L$) \cite{gegenwart2008quantum}. The ordering is then completely suppressed at the QCP, where only magnetic order parameter fluctuations are present and beyond this the system is paramagnetic, with a large FS (P$_L$). This scenario is generally referred to as a spin-density-wave QCP, with behavior consistent with the predictions from conventional Hertz-Millis models \cite{PhysRevB.14.1165Hertz, PhysRevB.48.7183Millis, Moriya}.

\subsection{Periodic Anderson model}

The periodic Anderson lattice model is the basic model for heavy fermion systems \cite{AndMod,HewsonHF,PColeman07Review}, which gives a microscopic picture for understanding the various possible ground states:
\begin{align}
\begin{split}
H=&\sum_{k\sigma}{\epsilon_kc_{k\sigma}^{+}c_{k\sigma}}+\epsilon_f\sum_{i\sigma}{n_{i\sigma}^{f}}+U\sum_{i}{n_{i\uparrow}^{f}n_{i\downarrow}^{f}}\\
&+ V \sum_{i\sigma}{\{f_{i\sigma}^{+}c_{i\sigma}+h.c.\}},
\end{split}
\end{align}
\noindent where $H_c=\sum_{k\sigma}{\epsilon_kc_{k\sigma}^{+}c_{k\sigma}}$ is the conduction electron energy, $H_f=\epsilon_f\sum_{i\sigma}{n_{i\sigma}^{f}}+U\sum_{i}{n_{i\uparrow}^{f}n_{i\downarrow}^{f}}$ contains the $f$ electron energy in the first term and the onsite Coulomb repulsion $U$ between electrons of opposite spins in the second term, and $H_{hyb} = V \sum_{i\sigma}{\{f_{i\sigma}^{+}c_{i\sigma}+h.c.\}}$ represents the hybridization between the $f$ and conduction electrons.

Many features of heavy fermion systems can be understood by considering a model with a single $f$ electron, the Anderson impurity model \cite{AndMod}. Here the $f$ electron can be either in the $ |f^{1}_{\uparrow}\rangle$ or the $ |f^{1}_{\downarrow}\rangle$ state \cite{PColeman07Review}. The energy cost of losing this electron to leave a vacant $|f^{0}\rangle$ state is $-\epsilon_f$, while $\epsilon_f+U$ is the energy for gaining an extra electron, to give a doubly occupied $|f^{2}\rangle$ state. In the heavy fermion regime, real charge fluctuations associated with  transitions from $ |f^{1}\rangle$ to the $|f^{0}\rangle$ or  $|f^{2}\rangle$  states are eliminated and the Ce valence remains integer. However, virtual processes render an effective antiferromagnetic interaction between the conduction electrons and local f moment in the form of the Kondo interaction $H_K=J\overrightarrow{\sigma}(0)\cdot \overrightarrow{S}_f$, with a coupling strength $J=V^2[\frac{1}{U+\epsilon_f}+\frac{1}{|\epsilon_f|}]$ \cite{SchrWol}. Similarly, the periodic Anderson model can be transformed into the Kondo lattice model given by Eq.~\ref{KLEq}.

Upon either increasing the hybridization $V$ between localized $f$ and conduction electrons or shifting the $f$ electron level $\epsilon_f$  closer to the Fermi energy E$_F$, valence fluctuations can no longer be neglected and the local moment approximation is not valid. The system smoothly evolves into the intermediate valence (mixed-valence) regime, where the Ce valence is not close to the trivalent integer limit \cite{PColeman07Review}. The electronic specific heat coefficient monotonically decreases with increasing $V$, due to a reduced effective electron mass and the  Curie-Weiss type behavior of the magnetic susceptibility is gradually suppressed with the disappearance of local moments.

\subsection{Valence Instability based on the generalized periodic Anderson lattice model}

\begin{figure}[t]
\includegraphics[angle=0,width=0.49\textwidth]{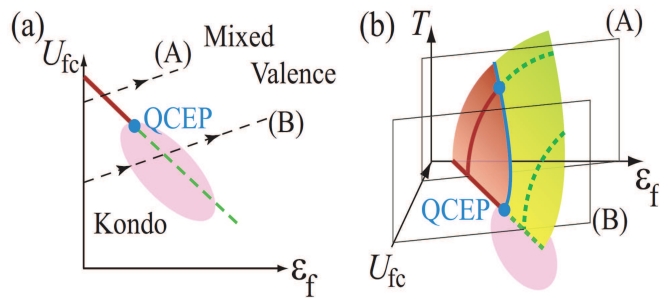}

\caption{\label{FigvalenceFL} (a) Schematic phase diagram of the ground states of the generalized  periodic Anderson lattice model in the U$_{fc}$-$\epsilon_f$ plane, where local Kondo and mixed valence regions are separated by a first-order valence transition (FOVT) line, which terminates at a quantum critical end point (QCEP). Lines (A) and (B) represent routes with a FOVT and continuous crossover behavior respectively. (b) Schematic $T-U_{fc}-\epsilon_f$ phase diagram derived from the generalized periodic Anderson lattice model, which extends (a) to finite temperatures. Adapted with permission from Ref.~\cite{SWatanabe11ValenceJPCM}. Copyright 2011 the Institute of Physics. }
\end{figure}

 In the Anderson lattice model, there is a smooth crossover from the heavy fermion to intermediate valence state. However, it has been proposed that the valence transitions or sharp valence crossovers observed in several heavy fermion systems can be understood from a generalized periodical Anderson lattice model \cite{Miyake00JPSJ}:
\begin{equation}
H=H_c+H_f+H_{hyb}+U_{fc}\sum_{i\sigma\sigma'}n_{i\sigma}^fn_{i\sigma'}^c,
\end{equation}

\noindent where the additional term with $U_{fc}$ represents the strength of the local Coulomb repulsion between the $f$ and conduction electrons. Using the mean-field approximation with the slave-boson formalism, it was demonstrated that as $\varepsilon_f$ moves towards the Fermi level, a drastic change of the Ce valence may take place at $\epsilon_f\sim E_F-U_{fc}$ \cite{Miyake00JPSJ}, where the $|f^{0}\rangle$ and $|f^{1}\rangle$ states are degenerate with enhanced valence fluctuations, assuming that $U_{fc}$ is of the order of the bandwidth of the conduction band \cite{Miyake00JPSJ,SWatanabe11ValenceJPCM}. As shown in Fig.~\ref{FigvalenceFL}(a) \cite{SWatanabe11ValenceJPCM}, there exists a first-order valence transition (FOVT) line in the schematic  phase diagram of the U$_{fc}$-$\epsilon_f$ plane, which terminates at a quantum critical end point (QCEP) and is connected to a sharp valence crossover region \cite{Miyake00JPSJ,SWatanabe11ValenceJPCM}. Adiabatic continuation between the Kondo and mixed-valence state can be achieved by avoiding the QCEP of the first-order transition line as in the liquid-gas case. In the temperature-U$_{fc}$-$\epsilon_f$ phase diagram shown in Fig.~\ref{FigvalenceFL}(b) \cite{SWatanabe11ValenceJPCM}, the surface cuts (A) and (B) respectively represent possible scenarios for the pressure induced behavior of Ce metal with a FOVT and Ce-based compounds with a sharp valence crossover.

At the critical valence transition, a few predictions have been made based on this model: \cite{Miyake02JPSJ,Holmes04PRB,Miyake00JPSJ} (1) a linear temperature dependence of the electrical resistivity; (2) an enhancement of the impurity scattering leading to a maximum in the residual resistivity; (3) a dramatic decrease of  electronic scattering, as reflected in the resistivity coefficient $A$, due to the sharp valence change between the Kondo and intermediate valence regimes. (4) In addition, a $d$-wave superconducting state may be stabilized by the significantly enhanced valence fluctuations, with $T_c$ reaching a maximum value.

The typical phase diagram of Ce-based heavy fermion systems may be summarized by Fig.~\ref{F1yuan} \cite{Yuan19122003}. Upon increasing the hybridization, e.g. by applying pressure, the long range antiferromagnetic order is eventually suppressed and a heavy fermion state develops. Near the AFM QCP, heavy fermion superconductivity is often observed. Upon further increasing the hybridization, the charge degrees of freedom may now play an important role and the system is driven into a mixed-valence state. Here the Coulomb interaction between $f$ and conduction electrons may lead to either a valence transition or crossover and superconductivity may also appear in proximity to the valence instability.

\section{Superconductivity near a magnetic instability}

\begin{table*}[t]
   \caption{Physical properties of Ce-based heavy fermion superconductors.}
   \label{tab1}
   \centering
   \tabcolsep=2pt
\begin{tabular}{c|cccccccc}\hline\hline
Compound       & Crystal   & $\gamma$         & $T_N$(p = 0)    & Propagation  & $p_c$    & $T_c$   & $H_{c2}$(//c) & $H_{c2}$($\perp$c)\\
               & Structure & [J/(mol~K$^2$)]  & (K)             & vector Q     & (GPa)    & (K)                       & (T)         & (T)   \\ \hline
CePd$_2$Si$_2$ & I4/mmm & 0.11 \cite{dhar_magnetism_1987} & 10 \cite{dhar_magnetism_1987} & (1/2, 1/2, 0) \cite{PhysRevB.29.2664} & 2.8 \cite{MathurQCP98Nature} & 0.5 \cite{MathurQCP98Nature} & 1.3 & 0.7 \cite{CePd2Si2Hc2} \\ \hline
CeNi$_2$Ge$_2$ & I4/mmm & 0.35 \cite{knopp_specific_1988} & - & - & - & 0.2 \cite{0953-8984-12-32-101} & \multicolumn{2}{c}{} \\ \hline
CeRh$_2$Si$_2$ & I4/mmm & 0.0228 \cite{PhysRevLett.78.3769} & 36 \cite{doi:10.1143/JPSJ.66.2260} & (1/2, 1/2, 0) & 1.06 \cite{0953-8984-14-21-102} & 0.4 \cite{0953-8984-14-21-102} & \multicolumn{2}{c}{0.28 \cite{0953-8984-14-21-102}} \\
               &        &                   &   & (1/2, 1/2, 1/2) ($T<$ 25~K) \cite{PhysRevB.29.2664} &      &     \\   \hline
CeCu$_2$Si$_2$ & I4/mmm & 1 \cite{FSteglich79PRL} & 0.8 \cite{PhysRevB.39.4726} & (0.215, 0.215, 0.53) \cite{stockert_magnetism_2008} & - & 0.6 \cite{PhysRevLett.52.469} & 2.4 & 2 \cite{PhysRevLett.52.469} \\ \hline
CeCu$_2$Ge$_2$ & I4/mmm & 0.1 \cite{DEBOER198791} & 4.15 \cite{DEBOER198791} & (0.28, 0.28, 0.54) \cite{CeCu2Ge2Neutron} & 10.1 \cite{Jaccard1992475} & 0.64 \cite{Jaccard1992475} & \multicolumn{2}{c}{2 \cite{Jaccard1992475}} \\ \hline
CeAu$_2$Si$_2$ & I4/mmm & 0.011 \cite{doi:10.1143/JPSJ.78.034714} & 10 \cite{PhysRevB.29.2664} & (0, 0, 0) \cite{PhysRevB.29.2664} & 22.5 \cite{PhysRevX.4.031055} & 2.5 \cite{PhysRevX.4.031055} & \multicolumn{2}{c}{9.2 \cite{PhysRevX.4.031055}} \\ \hline
CePt$_2$In$_7$ & I4/mmm & 0.34 \cite{PhysRevB.81.180507} & 5.5 \cite{PhysRevB.81.180507} & - & 3.5 \cite{PhysRevB.81.180507} & 2.1 \cite{PhysRevB.81.180507} & 15 \cite{PhysRevB.88.020503} & - \\ \hline
CeIn$_3$ & Pm$\bar{3}$m & 0.12 \cite{Sebastian12052009} & 10.23 \cite{PhysRevB.22.4379} & (1/2, 1/2, 1/2) \cite{Benoit1980293} & 2.65 \cite{knebel_quantum_2002} & 0.204 \cite{knebel_quantum_2002} & \multicolumn{2}{c}{0.45 \cite{PhysRevB.65.024425}} \\ \hline
CeCoIn$_5$ & P4/mmm & 0.29 \cite{CPetrovic01CeCoIn5} & - & - & - & 2.3 \cite{CPetrovic01CeCoIn5} & 5.2 & 11.5 \cite{1367-2630-13-11-113039} \\ \hline
CeRhIn$_5$ & P4/mmm & 0.4 \cite{PhysRevLett.84.4986} & 3.8 \cite{PhysRevLett.84.4986} & (1/2, 1/2, 0.297) \cite{PhysRevB.62.R14621} & 2.3 \cite{park_hidden_2006} & 2.258 \cite{doi:10.1143/JPSJ.77.114704} & 16.9 & 9.7 \cite{doi:10.1143/JPSJ.77.084708} \\ \hline
CeIrIn$_5$ & P4/mmm & 0.72 \cite{0295-5075-53-3-354} & - & - & - & 0.4 \cite{0295-5075-53-3-354} & 2.3 & 7 \cite{0295-5075-53-3-354}\\ \hline
Ce$_2$CoIn$_8$ & P4/mmm & 0.5 \cite{doi:10.1143/JPSJ.71.2836} & - & - & - & 0.4 \cite{doi:10.1143/JPSJ.71.2836} & 0.9 & 1.1 \cite{Hedo2004146} \\ \hline
Ce$_2$RhIn$_8$ & P4/mmm & 0.4 \cite{Thompson20015} & 2.8 \cite{Thompson20015} & (1/2, 1/2, 0) \cite{PhysRevB.64.020401} & 2.3 \cite{PhysRevB.67.020506} & 2 \cite{PhysRevB.67.020506} & \multicolumn{2}{c}{5.36 \cite{PhysRevB.67.020506}} \\ \hline
Ce$_2$PdIn$_8$ & P4/mmm & 1 \cite{Ce2PdIn8PRL} & 10 \cite{Ce2PdIn8PRL} & - & - & 0.68 \cite{Ce2PdIn8PRL} & 2.2 & 1.8 \cite{PhysRevB.84.140507} \\ \hline
CePt$_3$Si & P4mm & 0.39 \cite{CePt3Si2004} & 2.2 \cite{CePt3Si2004} & (0, 0, 1/2) \cite{doi:10.1143/JPSJS.81SB.SB006} & - & 0.75 \cite{CePt3Si2004} & 3.2 & 2.7 \cite{CePt3SiPress1}\\ \hline
CeCoGe$_3$ & I4mm & 0.111 \cite{CeCoGe31993} & 21 \cite{CeCoGe31993} & (0, 0, 1/2) ($T<$~8~K) \cite{CeCoGe3MS}  & 5.5 \cite{CeCoGe3SC} & 0.7 \cite{CeCoGe3SC} & 24 & 3.1 \cite{CeCoGe3Hc2b} \\  \hline
CeRhSi$_3$ & I4mm & 0.12 \cite{CeRhSi3Rep} & 1.6 \cite{CeRhSi3Rev} & ($\pm$0.215, 0, 0.5) \cite{CeRhSi3ND} & 2.36 \cite{CeRhSi3QCP} & 1.1 \cite{CeRhSi3Rev} & 30 & 7 \cite{CeRhSi3Hc2}\\\hline
CeIrSi$_3$ & I4mm & 0.125 \cite{CeRhSi3Rep} & 5 \cite{CeRhSi3Rep} & ($\pm$0.265, 0, 0.43) \cite{1742-6596-400-2-022003} & 2.5 \cite{CeIrSi3SC} & 1.6 \cite{CeIrSi3SC} & 45 & 9.5 \cite{CeIrSi3Hc2} \\ \hline
CeIrGe$_3$ & I4mm & 0.08 \cite{CeRhSi3Rep} & 8.7 \cite{CeRhSi3Rep} & - & 20 \cite{PhysRevB.81.140507} & 1.5 \cite{PhysRevB.81.140507} & $>$ 10 \cite{PhysRevB.81.140507} & - \\ \hline
\hline
\end{tabular}
\label{HFTable}
\end{table*}

\begin{figure*}[t]
\begin{center}
\includegraphics[angle=0,width=0.7\textwidth]{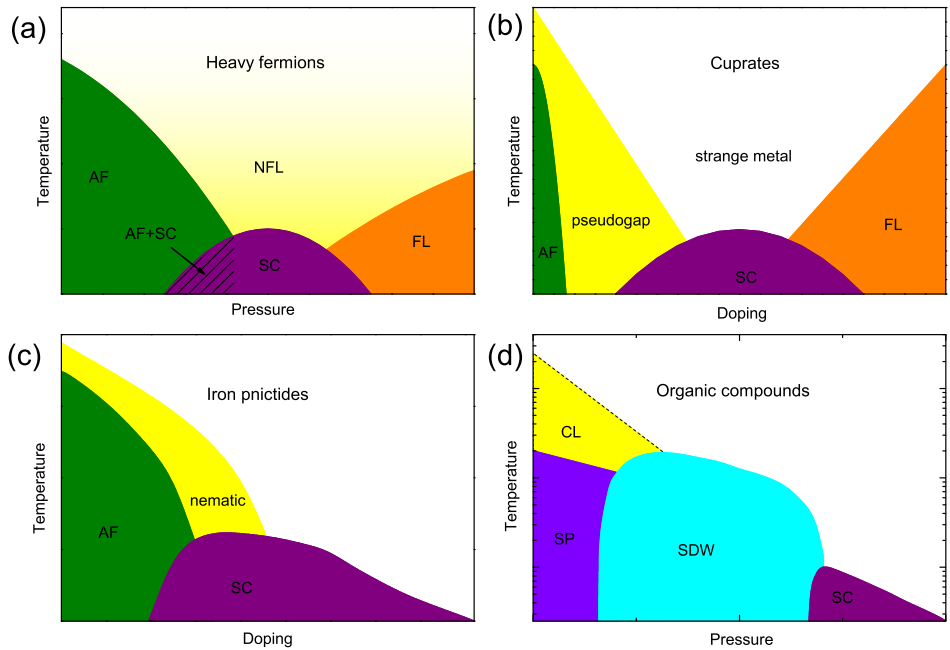}
\end{center}
\caption{\label{FigAFMSC} Phase diagrams for (a) heavy fermions (b) cuprates (c) iron pnictides and (d) organic compounds  \cite{JeromeOrg}. All of the diagrams show several similar features.}
\end{figure*}

As has been discussed previously, with increased coupling $J$, the strength of the Kondo interaction becomes greater than that of the RKKY interaction, which leads to the suppression of magnetic order at a QCP. In many Ce based heavy fermion systems, superconductivity emerges in the vicinity of this suppression. Examples of heavy fermion superconductors and their physical properties are displayed in Table.~\ref{HFTable}, where in some instances the system is superconducting at ambient pressure, while in other cases pressure must be applied for superconductivity to be realized. The phase diagrams of many Ce-based heavy fermion superconductors  are similar to that depicted in Fig.~\ref{FigAFMSC}(a). In the vicinity of an AFM QCP, non-Fermi liquid behavior in the normal state above $T_c$ is generally observed in the resistivity and specific heat, due to enhanced quantum critical spin fluctuations. Heavy Fermi liquid behavior is recovered upon crossing the QCP, below a characteristic Fermi liquid temperature $T_{FL}$, which increases with increasing pressure.

The close relationship between superconductivity and magnetism in heavy fermion systems is in stark contrast to conventional BCS superconductors, where the presence of magnetism is generally considered to be antagonistic to superconductivity. However, the presence of superconductivity on the border of a magnetic instability is also commonly found in other unconventional superconductors such as cuprate, iron-based and organic superconductors  \cite{JeromeOrg},  as shown in Fig.~\ref{FigAFMSC}. The nature of the antiferromagnetically ordered parent compound  varies, with Mott insulating, semi-metallic or intermetallic behavior observed, yet in all cases the superconductivity is in the vicinity of magnetism. Even though the $T_c$ of heavy fermion superconductors is usually much lower than cuprate or iron-based superconductors, the ratio $T_c/T_F$  is much higher than BCS superconductors  \cite{FSteglich79PRL}. This is because the effective Fermi temperature $T_F$ is much smaller, on the order of the coherence temperature, compared to $\sim10^4$~K in simple metals. In this sense, heavy fermion superconductors may also be regarded as analogous to the high $T_c$ materials.

In a simple phenomenological model \cite{DPines07Nature}, the spin interaction between quasiparticles can be given by
\begin{equation}
V_{ind}(r,t)=-s\cdot s'g_m^2 \chi_m(r,t),
\end{equation}

\noindent where $g_m$ characterizes the coupling strength of the spins $s$ and $s'$, while $\chi_m(r,t)$ is the non-local magnetic susceptibility. On the border of  AFM long-range order, $\chi_m(r,t)$ is expected to be dramatically enhanced due to quantum fluctuations and to oscillate in real space, giving the attractive interaction necessary for the formation of Cooper pairs in spin-singlet states with non-zero angular momentum, such as $d$-wave superconductivity. However, experimentally characterizing the pairing symmetry of many heavy fermion superconductors remains a challenge. In particular, for those systems where superconductivity requires high pressures and very low temperatures, the range of experimental techniques available for measuring the pairing symmetry is greatly reduced. Also in noncentrosymmetric heavy fermion superconductors \cite{CePt3Si2004}, the effects of inversion symmetry breaking and antisymmetric spin-orbit coupling need to be taken into account, which will generally lead to very different pairing states with a mixture of singlet and triplet components \cite{NCSGorkov}.

While the presence of superconductivity on the border of magnetism is well established for heavy fermion superconductors, the detailed relationship between superconductivity, AFM order and quantum criticality in different systems is the subject of much ongoing interest. There are particular questions about the degree of competition or coexistence between magnetism and superconductivity in these systems and the extent to which magnetism persists within the superconducting dome. In addition, the role of quantum criticality in driving superconductivity is not well understood, particularly in those systems where there is evidence for a first-order transition between the AFM and paramagnetic states and a lack of evidence for enhanced critical fluctuations. In the remainder of this section, we describe in detail the properties of several heavy fermion superconductors, with a particular focus on the relationship between superconductivity and magnetic order and the nature of the superconducting phases which emerge.

\subsection{ Superconductivity near an antiferromagnetic QCP}

\begin{figure}[t]
\includegraphics[angle=0,width=0.49\textwidth]{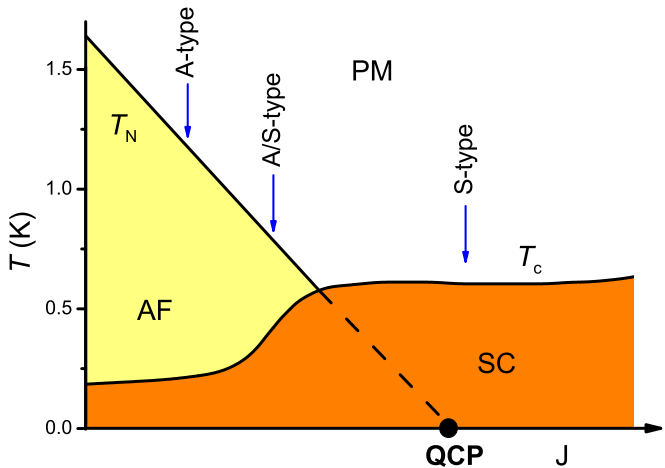}

\caption{\label{CCSPhasediag} Schematic phase diagram of CeCu$_2$Si$_2$ \cite{stockert_magnetically_2011}, where the parameter $J$ can be tuned either by applying pressure, or by small changes to the composition. }
\end{figure}

CeCu$_2$Si$_2$ was the first heavy fermion system \cite{FSteglich79PRL} found to be superconducting and  the ground state can be categorized as $A$-, $A/S$- or $S$-type, where $A$ and $S$ denote antiferromagnetism and superconductivity \cite{CeCu2Si2AS} respectively. A schematic phase diagram is displayed in Fig.~\ref{CCSPhasediag} \cite{stockert_magnetically_2011}, where at ambient pressure the hybridization tuning parameter $J$ depends on the precise Cu composition \cite{CCSFlux} and $J$ is increased by applying pressure. Upon tuning $J$ through a QCP, the magnetism is entirely suppressed and only a superconducting transition is observed. The $A$-type samples order antiferromagnetically at $T_N$ and may also display superconductivity at a much lower $T_c$, while $S$-type samples display only superconductivity. On the other hand in $A/S$-type samples, both antiferromagnetic and superconducting transitions are detected with similar values of $T_N$ and $T_c$. The antiferromagnetic phase consists of incommensurate SDW order with a propagation vector $\mathbf{q}~=~(0.215,0.215,0.530)$ and a magnetic moment of $\sim$ 0.1 $\mu_B$/Ce \cite{stockert_magnetism_2008}. Muon spin relaxation ($\mu$SR) measurements \cite{CeCu2Si2muSR1,CeCu2Si2muSR2} of $A/S$-type samples show that upon entering the superconducting state, the magnetic volume fraction decreases, indicating there is competition between magnetism and superconductivity and a separation of the two phases on a microscopic level.

\begin{figure*}[t]
\begin{center}
\includegraphics[angle=0,width=0.75\textwidth]{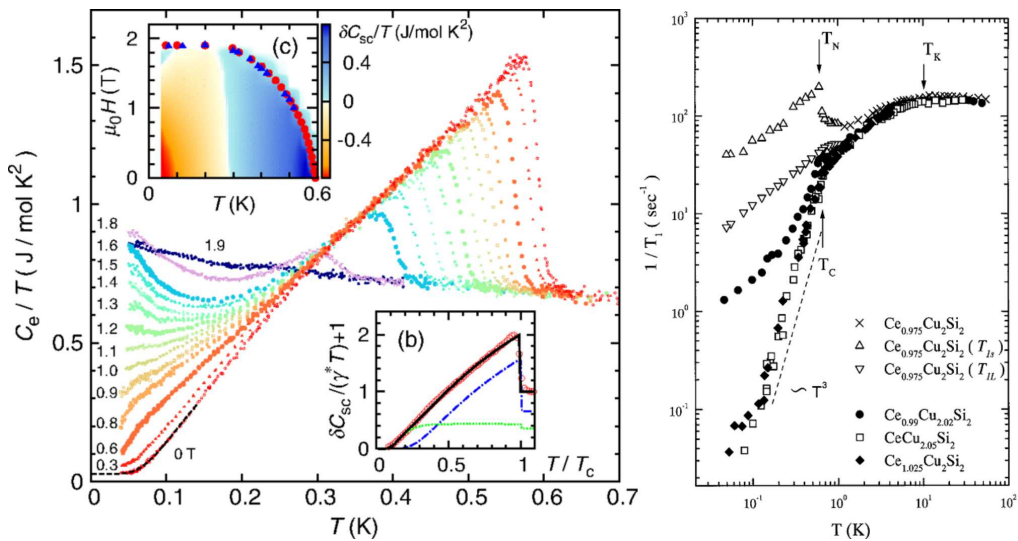}
\end{center}
\caption{ Specific heat (left) and NQR (right) measurements of superconducting CeCu$_2$Si$_2$ to low temperatures. The insets show the temperature-field phase diagram and the fitting of the data with a two gap, nodeless model.  Adapted with permission from Ref.~\cite{PhysRevLett.112.067002}, copyright 2014 the American Physical Society and Ref.~\cite{PhysRevLett.82.5353}, copyright 1999 the American Physical Society.}\label{CeCuSiorderparameter}
\end{figure*}

Despite experimental efforts, the superconducting gap structure and pairing symmetry, which are closely related to the pairing mechanism, still remain controversial and are the subject of debate. It was generally believed that the gap symmetry of ${\mathrm{CeCu}_2\mathrm{Si}_2}$ should be even parity $d$-wave with line nodes. The even parity is a consequence of spin-singlet superconductivity, which was inferred from both the evidence for Pauli paramagnetic limiting of the upper critical field  $H_{c2}$ \cite{CCSPauli} and a decrease of the NMR Knight shift below $T_c$ \cite{CCSKS}. As shown in Fig.~\ref{CeCuSiorderparameter}(b), the presence of line nodes was supported by both a $T^3$ dependence and the absence of a coherence peak in the spin lattice relaxation rate $1/T_1$ from NQR measurements \cite{CCSNMR1,PhysRevLett.82.5353}. Inelastic neutron scattering and the subsequent theoretical analysis suggested $d_{x^2 - y^2}$ symmetry for ${\mathrm{CeCu}_2\mathrm{Si}_2}$ \cite{stockert_magnetism_2008, PhysRevLett.101.187001}, while the four-fold oscillations observed in the angular dependence of $H_{c2}$ are consistent with $d_{xy}$ symmetry \cite{PhysRevLett.106.207001}. However, recent low-temperature specific heat measurements of ${\mathrm{CeCu}_2\mathrm{Si}_2}$ exhibit an exponential temperature dependence (Fig.~\ref{CeCuSiorderparameter}(a)), characteristic of fully gapped superconductivity \cite{PhysRevLett.112.067002}.  In addition, both the recent specific heat \cite{PhysRevLett.112.067002} and scanning tunneling spectroscopy \cite{CCSSTM} measurements  also indicate the presence of multiple superconducting gaps . An alternative loop nodal $s\pm$ pairing symmetry was proposed \cite{PhysRevLett.114.147003}. However, it is difficult to explain the results of inelastic neutron scattering measurements \cite{stockert_magnetically_2011} using either the isotropic two-band $s$-wave or loop nodal $s\pm$ scenarios, since they do not provide the deduced sign change of the superconducting gap between regions of the Fermi surface spanned by the antiferromagnetic ordering wave vector. Recent  penetration depth measurements also support the presence of fully gapped superconductivity and we propose a fully-gapped $d+d$ band-mixing pairing state, which changes sign over the Fermi surface and therefore appears to be consistent with the literature results \cite{pang2016evidence}.

A number of additional heavy fermion superconductors have also been discovered. CeIn$_3$ is antiferromagnetic, with $T_N\sim$ 10~K at ambient pressure, but the magnetism is entirely suppressed by applying a pressure of around 2.6~GPa \cite{knebel_quantum_2002}. Superconductivity emerges near the region where the suppression occurs, with a maximum $T_c$ less than 0.3~K. NQR measurements \cite{CeIn3NMR} in the superconducting state when magnetism has been suppressed show a lack of a coherence peak and a $T^3$ dependence of $1/T_1$. These observations  indicate that there are  line nodes in the superconducting gap. Introducing a buffer layer $M$In$_2$ ($M$~=~Co, Rh or Ir) into the superconducting block CeIn$_3$ lead to the discovery of the so called 115 family \cite{ThompsonReview1, ThompsonReview2}, which have become an important family of heavy fermion superconductors. For instance, CeCoIn$_5$ \cite{CPetrovic01CeCoIn5, SidorovCeCoIn502PRL} and CeRhIn$_5$ \cite{park_hidden_2006, CeRhIn5ParkNJPhys} have the highest $T_c$ of Ce-based heavy fermion materials, either at ambient or applied pressure, which is generally ascribed to the quasi two-dimensional nature of the spin fluctuations. Further analogous compounds with the formula Ce$_n$M$_m$In$_{3n+2m}$ have also been synthesized, such as Ce$_3$PdIn$_{11}$ \cite{Tursina20137, Ce3PdIn11SciRep15}, Ce$_2$PdIn$_8$ \cite{Ce2PdIn8PRL}, and CePt$_2$In$_7$ \cite{PhysRevB.81.180507, PhysRevB.88.020503}. 

${\mathrm{CePd}}_{2}{\mathrm{Si}}_{2}$ is also a  heavy Fermion compound with an enhanced specific heat coefficient $\gamma \simeq$ 110~mJ/(mol~K$^2$) \cite{dhar_magnetism_1987}. ${\mathrm{CePd}}_{2}{\mathrm{Si}}_{2}$ enters an antiferromagnetic state below $T_N$ = 10~K \cite{PhysRevB.29.2664}, very close to its Kondo temperature $T_K$ = 10~K \cite{PhysRevB.39.4164}. The ordered moment is just 0.62$\mu_B$, with $\mathbf{q} = (\frac{1}{2}, \frac{1}{2}, 0)$, as confirmed by neutron scattering measurements at ambient pressure \cite{PhysRevB.29.2664}. With increasing pressure, $T_N$ monotonically decreases and drops below 1~K at around 2.6~GPa \cite{MathurQCP98Nature}, while neutron diffraction measurements confirm a decrease of the ordered moment \cite{PhysRevB.71.064404}. A linear extrapolation of the $T_N-p$ curve to zero temperature gives a critical pressure of $p_c$~=~2.8~GPa. Slightly below $p_c$, superconductivity emerges, which reaches a maximum transition temperature of $T_c$~=~0.5~K at $p_c$ \cite{MathurQCP98Nature}. Around $p_c$, the electrical resistivity shows non-Fermi liquid behavior up to temperatures greater than 20~K, with an almost linear relation $\Delta\rho \propto T^{1.2}$.  The magnetic transition at $p_c$ is generally believed to be second-order, while a first-order transition can not be entirely excluded due to the limitation of pressure \cite{PhysRevB.71.064404}. The angular dependent de Haas-van Alphen (dHvA) effect measurements \cite{sheikin_haasvan_2004} are consistent with band structure calculations with the assumption of itinerant $4f$ electrons in the antiferromagnetic state, suggesting there is an itinerant SDW type quantum critical point at $p_c$ in ${\mathrm{CePd}}_{2}{\mathrm{Si}}_{2}$.

\begin{figure}[t]
\includegraphics[width=8.3cm, height=6.26cm]{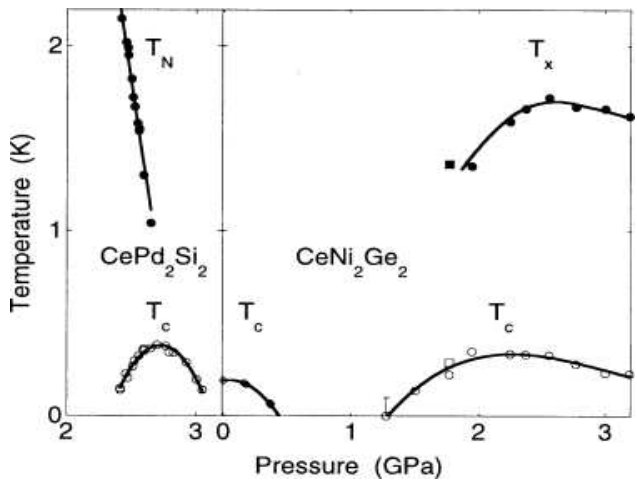}

\caption{Temperature-pressure phase diagram of ${\mathrm{CePd}}_{2}{\mathrm{Si}}_{2}$ (left) and ${\mathrm{CeNi}}_{2}{\mathrm{Ge}}_{2}$ (right). The pressure dependence of the ordering temperature $T_N$ and superconducting transition $T_c$ are shown. Reprinted with permission from Ref.~\cite{0953-8984-12-32-101}. Copyright 2000 the Institute of Physics. }\label{F4Grosche}
\end{figure}

The isoelectronic compound ${\mathrm{CeNi}}_{2}{\mathrm{Ge}}_{2}$ does not magnetically order at ambient pressure \cite{knopp_specific_1988}. However, $C/T$ exhibits a logarithmic divergence at low temperatures with a large Sommerfeld coefficient $\gamma \simeq 350$~mJ/(mol~K$^2$) \cite{PhysRevLett.82.1293, knopp_specific_1988}. The temperature dependence of the resistivity also shows  non-Fermi liquid behavior, with a reported exponent $n$  between 1.1 and 1.5 \cite{PhysRevLett.82.1293, 0953-8984-12-32-101}, where higher quality samples generally have a smaller $n$. It has been proposed that ${\mathrm{CeNi}}_{2}{\mathrm{Ge}}_{2}$ is located near a three dimensional antiferromagnetic QCP \cite{1468-6996-8-5-A17}, and the system can be driven to a QCP by doping the Ni site with about 6.5\% Pd \cite{CNGxc,0953-8984-27-1-015602}. Evidence for a superconducting transition also appears below 0.2~K in resistivity measurements, although zero-resistance has not been reached, and NQR measurements were reported to show a slight decrease in $1/T_1T$, probably caused by this filamentary superconductivity \cite{doi:10.1143/JPSJ.75.043702}. No bulk superconductivity is reported so far and the existence of superconductivity still requires further confirmation. ${\mathrm{CePd}}_{2}{\mathrm{Si}}_{2}$ could be considered to correspond to ${\mathrm{CeNi}}_{2}{\mathrm{Ge}}_{2}$ at negative pressure, while ${\mathrm{CeNi}}_{2}{\mathrm{Ge}}_{2}$ is situated very close to  a magnetic instability at ambient pressure. An interesting feature of ${\mathrm{CeNi}}_{2}{\mathrm{Ge}}_{2}$ is the occurrence of a second superconducting region under pressure, ranging from 1.5 to 6.5~GPa \cite{0953-8984-12-32-101}. The phase diagram (Fig.~\ref{F4Grosche}) encompassing both ${\mathrm{CeNi}}_{2}{\mathrm{Ge}}_{2}$ and ${\mathrm{CePd}}_{2}{\mathrm{Si}}_{2}$ is reminiscent of the phase diagram of ${\mathrm{CeCu}}_{2}{\mathrm{(Si,Ge)}}_{2}$ \cite{Yuan19122003} discussed in the next section.

\subsection{Superconductivity near a possible first-order magnetic quantum phase transition}

\begin{figure}[t]

\includegraphics[angle=0,width=0.48\textwidth]{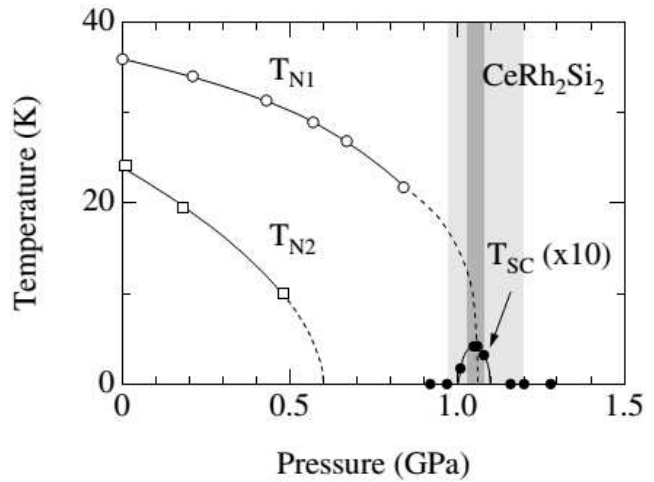}
\caption{Temperature-pressure phase diagram of  ${\mathrm{CeRh}}_{2}{\mathrm{Si}}_{2}$.  The ordering temperatures $T_{N1}$ and $T_{N2}$ are suppressed by pressure, with superconductivity being observed near the disappearance of $T_{N1}$.  Reprinted with permission from Ref.~\cite{0953-8984-14-21-102}. Copyright 2002 the Institute of Physics.} \label{F1Araki}
\end{figure}

${\mathrm{CeRh}}_{2}{\mathrm{Si}}_{2}$ is another heavy fermion compound with the same tetragonal ${\mathrm{ThCr}}_{2}{\mathrm{Si}}_{2}$-type crystal structure as CeCu$_2$Si$_2$. ${\mathrm{CeRh}}_{2}{\mathrm{Si}}_{2}$ undergoes an antiferromagnetic transition at $T_{N1}$~=~36~K to a state AF1 with propagation vector $\mathbf{q_1} = (\frac{1}{2},\frac{1}{2},0)$, before entering a second antiferromagnetic state (AF2) at $T_{N2}$~=~25~K via a first-order transition with an additional propagation vector $\mathbf{q_2} = (\frac{1}{2},\frac{1}{2},\frac{1}{2})$ \cite{PhysRevB.57.7442, PhysRevB.61.4167}. As shown in Fig.~\ref{F1Araki} \cite{0953-8984-14-21-102}, both AF1 and AF2 can be completely suppressed by pressure, at $p_{c} \simeq$ 1.06~GPa and $p_{c}^{'} \simeq$ 0.6~GPa respectively. Unconventional superconductivity is observed around $p_c$ with a maximum $T_c \simeq$ 0.4~K \cite{PhysRevB.53.8241,0953-8984-14-21-102}.

The dHvA measurements under pressure indicate that the Fermi surface topology remains unchanged below $p_c$ \cite{PhysRevB.64.224417}. Some of the branches observed at low pressure are consistent with the band structure calculations of ${\mathrm{LaRh}}_{2}{\mathrm{Si}}_{2}$, indicating there are localized $4f$~electrons, which do not contribute to the volume of the Fermi surface. Note that some branches either  appear or disappear at 0.5~GPa, due to the change of magnetic structure  inside the antiferromagnetic state. When the pressure is increased through $p_c$, the topology of the Fermi surface undergoes a sudden change, suggesting that the localized $4f$ electrons may become itinerant. However, thermal expansion measurements show a clear discontinuity in the cell volume, which suggests that the transition from antiferromagnetism to paramagnetism at $p_c$ is probably first-order\cite{0953-8984-20-1-015203}. Although the $A$ coefficient and residual resistivity $\rho_0$ reach a maximum around $p_c$, a $T^2$ dependence of the resistivity holds across the whole pressure range, in contrast to the non-Fermi liquid behavior generally observed around a QCP \cite{0953-8984-14-21-102}. The value of $\gamma$ only increases to 80~mJ/(mol~K$^2$) at $p_c$ before decreasing smoothly above $p_c$ \cite{PhysRevLett.78.3769}. This corresponds to only a moderate enhancement of the effective mass, indicating that a significant enhancement of critical fluctuations around $p_c$ is probably avoided  in ${\mathrm{CeRh}}_{2}{\mathrm{Si}}_{2}$, due to the first-order nature of the AFM to paramagnetic transition.

Measurements of ${\mathrm{CeRh}}_{2}{\mathrm{Si}}_{2}$ under pressure can be compared to those in magnetic fields, using techniques such as torque, magnetostriction and transport measurements  \cite{PhysRevB.81.094403}. The  field-temperature phase diagram demonstrates that both AF1 and AF2 are gradually suppressed with increasing magnetic field and merge into a tetra-critical point $T_{TCP}$, with a third phase AF3 emerging in a narrow  field range between 25.7 and 26~T. Above the critical magnetic field $H_c \sim $ 26.0~T, the system enters the polarized paramagnetic state. The resistivity coefficient $A$ and residual resistivity show similar behavior at $H_c$ as that under pressure, while the abrupt change in the resistivity at 26~T suggests a probable Fermi surface reconstruction at $H_c$. Whether this reconstruction corresponds to a Kondo-breakdown QCP or first-order transition remains unresolved. It has also been suggested that ${\mathrm{CeRh}}_{2}{\mathrm{Si}}_{2}$ could be pushed to a valence critical point by applying a magnetic field at pressures greater than $p_c$, which may explain recent thermoelectric power (TEP) measurements \cite{PhysRevB.64.224417}.

\subsection{Noncentrosymmetric heavy fermion superconductors}

\begin{figure*}[t]
\begin{center}
\includegraphics[angle=0,width=0.65\textwidth]{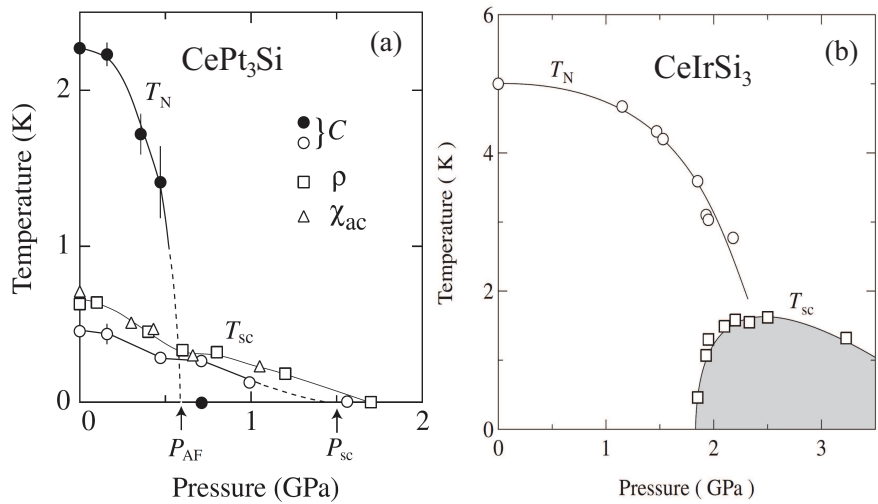}
\end{center}
\caption{Temperature-pressure phase diagrams of noncentrosymmetric (a) CePt$_3$Si and (b) CeIrSi$_3$. Adapted with permission from Ref.~\cite{CePt3SiPress2}, copyright 2005 the Physical Society of Japan and Ref.~\cite{CeIrSi3SC},  copyright 2006 the Physical Society of Japan respectively. \label{NCSPhase}}
\end{figure*}

Another widely studied class of heavy fermion superconductors are those with noncentrosymmetric crystal structures. One consequence of the lack of inversion symmetry is that in the presence of  antisymmetric spin orbit coupling, the superconducting state is neither purely singlet or triplet, but a mixture of the two \cite{NCSGorkov}. CePt$_3$Si is one such compound which orders antiferromagnetically at $T_N~=~2.2$~K and becomes superconducting at $T_c~=~0.75$~K \cite{CePt3Si2004}. Strong hybridization between the $4f$ and conduction electrons is evidenced by the enhanced value of $\gamma$~=~390~mJ/(mol~K$^2$) and a small ordered moment of 0.16~$\mu_B$/Ce \cite{CePt3SiMagStruc}, which is significantly reduced below the value expected from the crystalline-electric-field (CEF) scheme \cite{CePt3SiCEF}. Upon applying pressure, $T_N$ decreases and is quickly suppressed around $0.5~-~0.6$~GPa (Fig.~\ref{NCSPhase}(a))      \cite{CePt3SiPress1,CePt3SiPress2}. However, unlike many other heavy fermion superconductors which show a superconducting dome with a maximum of $T_c$ in close proximity to the suppression of magnetic order, the $T_c$ of CePt$_3$Si is largest at ambient pressure and decreases upon increasing the pressure before disappearing around 1.5~GPa. These results indicate a lack of competition between the superconductivity and magnetism in CePt$_3$Si. This is further supported by $\mu$SR measurements \cite{CePt3SiMuSR}, which indicate a microscopic coexistence between the phases. In addition, ac heat capacity measurements under pressure \cite{CePt3SiPress2}, show no increase in the heat capacity jump at $T_c$ ($\Delta C_{ac}/C_{ac}(T_c)$)  when $T_N$ is suppressed by pressure.

A larger family of noncentrosymmetric heavy fermions are Ce$TX_3$ ($T$~=~transition metal, $X$~=~Si or Ge), which crystallize in the tetragonal BaNiSn$_3$ structure with space group $I4mm$. Some of these compounds such as CeRhGe$_3$, CeIrGe$_3$ and CePtSi$_3$ behave as localized antiferromagnets \cite{CeRhSi3Rep,CeRhGe32012,CePtSi3Mag}. Of these,  CeIrGe$_3$ displays superconductivity at  high pressures, greater than 20~GPa \cite{CeIrGe3SC}, while the other two do not show any evidence of superconductivity up to at least 8~GPa \cite{CeTX32008,CePtSi3Press}. Other isostructural compounds such as CeCoSi$_3$ and CeRuSi$_3$ are non-magnetic intermediate valence compounds \cite{CeTX32008,CeRuSi3IV} where the local moments are entirely quenched and have large values of $T_K$. The main focus of interest have been from compounds in between these two extremes, namely CeRhSi$_3$, CeIrSi$_3$ and CeCoGe$_3$. Both CeRhSi$_3$ and CeIrSi$_3$ order antiferromagnetically at $T_N~=~1.6$~K and $5$~K respectively \cite{CeRhSi3Rep,CeRhSi3Rev}. The magnetic order of both compounds is of the incommensurate spin density wave type \cite{CeRhSi3ND,CeIrSi3MagStruc}, with the magnetic moments orientated in the $ab$~plane and dHvA measurements \cite{CeRhSi3FS1,CeRhSi3FS2,CeIrSi3Itin} show significant differences to the respective La based analogues, indicating the $4f$ electrons are itinerant. Furthermore, upon applying pressure up to nearly 3~GPa, no drastic change in the Fermi surface of CeRhSi$_3$ was observed and the effective mass steadily decreases with pressure. The temperature-pressure phase diagrams are similar to many other heavy fermion superconductors. While $T_N$ of CeRhSi$_3$ \cite{CeRhSi3SC,CeRhSi3Rev} initially slightly increases with applied pressure, it is subsequently suppressed to zero temperature around $p_c~=~2.36$~GPa, while a superconducting dome emerges for pressures greater than $1.2$~GPa, with $T_c$ reaching a maximum slightly above $p_c$ \cite{CeRhSi3Hc2}. Evidence for the presence of a QCP at $p_c$ was inferred from $\mu$SR measurements, where $T_N$ and the ordered moment of CeRhSi$_3$ were both reported to be suppressed to zero at a second-order transition at $p_c$ \cite{CeRhSi3QCP}. However, in this region no increase of the $A$ coefficient or residual resistivity is observed, indicating there is little enhancement of spin fluctuations \cite{CeRhSi3NoQCP}. For CeIrSi$_3$ \cite{CeIrSi3SC} (Fig.~\ref{NCSPhase}(b)), $T_N$ continuously decreases with increasing pressure and superconductivity is observed above 1.8~GPa, with a maximum $T_c$ of around 1.6~K at 2.5~GPa, which is approximately where $T_N$ intersects the superconducting dome.

Different behavior is observed for CeCoGe$_3$ \cite{CeCoGe31993,CeCoGe32005} which has a higher ordering temperature of $T_N~=~21$~K. At ambient pressure the $4f$ electrons are localized \cite{CeCoGe3FS,CeCoGe3FSHF} and the magnetic structure is commensurate, consisting of moments ordered along the $c$~axis \cite{CeCoGe3SCND,CeCoGe3PND,CeCoGe3MS}. The temperature-pressure phase diagram is more complex, with multiple pressure induced magnetic phases and $T_N$ undergoes a series of step-like decreases \cite{CeCoGe3LP,CeCoGe3HP}. At around $p_c~=~5.5$~GPa, magnetism is suppressed while a broad superconducting dome is observed \cite{CeCoGe3SC,CeCoGe3HP} from 4.3~GPa, to at least 7.1~GPa. Therefore the properties of the Ce$TX_3$ have generally been considered to be consistent with the Doniach phase diagram \cite{CeTX32008}, where the system can be tuned from localized magnetism with weak hybridization to intermediate valence behavior with strong hybridization by either elemental substitution or pressure. The effect of chemical pressure may not be sufficient to explain the behavior of all Ce$TX_3$ compounds and in particular, for different transition metals $T$, other changes to the electronic properties may need to be taken into account.

In both the crystal structures of CePt$_3$Si and Ce$TX_3$, inversion symmetry is absent due to the lack of a mirror plane along the $c$~direction. This gives rise to a Rashba type antisymmetric spin-orbit coupling which in general should lead to mixed parity pairing. Indeed there is evidence for unconventional superconductivity in both CePt$_3$Si \cite{CePt3SinodesCond,CePt3SiBonalde2004} and  CeIrSi$_3$ \cite{CeIrSi3NMR1} from measurements of the superconducting gap structure, which indicate the presence of line nodes in the gap. A particular prediction for the effects of antisymmetric spin-orbit coupling for these crystal structures is an anisotropic spin susceptibility below $T_c$, which is expected to remain unchanged from the normal state value for fields along the $c$~axis, while reaching half the normal state value for fields in the $ab$ plane \cite{FrigeriSusc}. This may be probed either by measuring the NMR Knight shift or the anisotropy of the upper critical field $H_{c2}$. In the case of the Ce$TX_3$ superconductors, $H_{c2}$ is highly anisotropic \cite{CeRhSi3Hc2,CeIrSi3Hc2,CeCoGe3Hc2b}, and in CeRhSi$_3$ it reaches around 30~T for fields along the $c$~axis, but just 7~T in the $ab$~plane, which is consistent with the effects of  antisymmetric spin-orbit coupling. In addition, the Knight shift of CeIrSi$_3$ \cite{CeIrSi3NMR2} is also qualitatively consistent, although the reduction below $T_c$ reaches just 90\% of the normal state value, unlike the 50\% predicted by the simple model. On the other hand, the $H_{c2}$ of CePt$_3$Si only shows a small anisotropy \cite{CePt3SiNMR2} and the Knight shift appears constant below $T_c$ in all directions \cite{CePt3SiNMR2}, which remains difficult to account for within this scenario.

\section{Superconductivity near a Charge Instability}

As described in the preceding section, superconductivity in Ce-based heavy fermion metals usually appears while suppressing the AFM order by applying pressure, showing a maximum $T_c$ near the QCP. In this case, it is generally believed that superconducting electrons are paired through critical spin fluctuations. However, the first heavy fermion superconductor CeCu$_2$Si$_2$, and its sister compounds CeCu$_2$Ge$_2$ exhibit highly unusual behavior under pressure, with superconductivity extending to very high pressure \cite{THOMAS1993303, Jaccard19991} and a large enhancement of $T_c$ far away from the AFM QCP which is located at ambient pressure for CeCu$_2$Si$_2$ and $p_{c1}\simeq 11.5$~GPa for CeCu$_2$Ge$_2$ respectively. Recently, similar phenomena were also shown to occur in CeAu$_2$Si$_2$ under pressure, where $p_{c1}\simeq 22.5$~GPa \cite{PhysRevX.4.031055}. These findings remain a puzzle and challenge the general view that heavy fermion superconductivity only arises via spin fluctuation mediated pairing.

 Several theoretical scenarios have been proposed to account for the anomalous properties in CeCu$_2$Si$_2$ and CeCu$_2$Ge$_2$. For instance, Thomas $\textit{et al}$ attributed the enhancement of $T_c$ to pressure-induced topological changes of the renormalized heavy bands \cite{thomas1996}. Onishi and Miyake studied a generalized Anderson lattice model and found that the local Coulomb repulsion $U_{fc}$ between the $f$ and the conduction electrons can lead to a valence transition and proposed that superconductivity in CeCu$_2$Si$_2$ and CeCu$_2$Ge$_2$ under pressure is mediated by valence fluctuations \cite{Miyake00JPSJ}. A three dimensional periodic Anderson model was also investigated by Ikeda using third-order perturbation theory \cite{doi:10.1143/JPSJ.71.1126}. It was argued that the dominant interaction is the on-site repulsion $U$ between $f$ electrons and as a result it is difficult to realize superconductivity arising from $U_{cf}$ rather than from $U$. Instead, he suggested that the $T_c$ enhancement arises from a dramatic change of the Fermi surface topology or a large momentum-independent mass enhancement.

\begin{figure}[t]
\includegraphics[angle=0,width=0.48\textwidth]{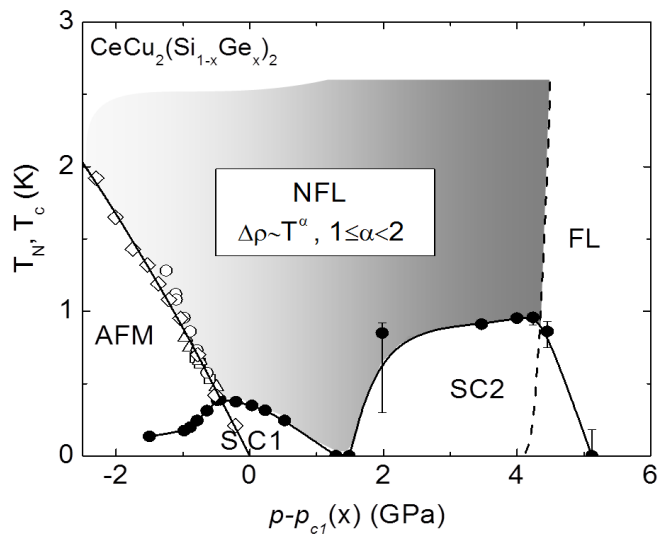}

\caption{ Temperature-pressure phase diagram of CeCu$_2$(Si$_{1-x}$Ge$_x$)$_2$, showing two superconducting domes under pressure, where $p_{c1}(x)$ is the critical pressure at the AFM QCP for a given Ge doping $x$. The ordering temperature $T_N$ is shown by the diamonds ($x=0.25$), open circles ($x=0.1$), triangles ($x=0.05$) and squares ($x=0.01$), while the superconducting transition $T_c$ is shown by the closed circles  ($x=0.1$).  Adapted with permission from Ref.~\cite{YuanPRL}. Copyright 2006 the American Physical Society.}\label{F1yuanPRL}
\end{figure}

Further progress towards understanding the behavior of CeCu$_2$Si$_2$ and CeCu$_2$Ge$_2$ came from the discovery that the superconductivity in the phase diagram comes from the merger of two different superconducting domes \cite{Yuanfirst, Yuan19122003, YuanPRL}. In CeCu$_2$Si$_2$, the partial substitution of Ge for Si has two main effects \cite{1367-2630-6-1-132}. Firstly, the crystal lattice is expanded and therefore magnetic order is stabilized. Secondly, such a substitution introduces additional disorder and reduces the mean free path, which may suppress superconductivity. Upon pressurizing CeCu$_2$(Si$_{1-x}$Ge$_x$)$_2$, the system corresponds to the pure compound CeCu$_2$Si$_2$ but with enhanced scattering due to disorder. In fact, it was found that $T_N$ for samples with various Ge-content overlap well after taking into account of the shift of the chemical pressure (Fig.~\ref{F1yuanPRL}), suggesting that the magnetic order is robust against disorder and $T_N$  is primarily determined by the lattice parameters. On the other hand, superconductivity in pressurized CeCu$_2$(Si$_{1-x}$Ge$_x$)$_2$ is strongly affected by the substitution of Ge.

 In Fig.~\ref{F1yuanPRL} \cite{YuanPRL}, the $T-p$ phase diagram of CeCu$_2$(Si$_{1-x}$Ge$_x$)$_2$ is displayed where the zero of the pressure axis corresponds to the AFM QCP at $p_{c1}(x)$ for each Ge-concentration. It is noted that both CeCu$_2$Si$_2$ and CeCu$_2$Ge$_2$ display a similar phase diagram, showing a continuous superconducting regime over a wide pressure range \cite{THOMAS1993303, Jaccard19991}. By partially substituting Si with Ge ($x=0.1$), $T_c$ is lowered and the superconducting regime is separated into two disconnected superconducting domes, one sitting near an AFM QCP and the other one taking place at higher pressure with a maximum $T_c$, near to a possible valence critical end point \cite{Yuan19122003,YuanPRL}. Upon further increasing the Ge-content, no superconductivity is observed under pressure. These findings at low pressures demonstrate that CeCu$_2$Si$_2$ resembles several other Ce-based heavy fermion systems, supporting the common feature of magnetically mediated superconductivity in heavy fermion superconductors \cite{MathurQCP98Nature}. In contrast to many other systems, a new superconducting state emerges at elevated pressures in CeCu$_2$Si$_2$ and some other isotructural compounds, which has been taken to be the first example of valence-fluctuation mediated superconductivity and is the focus of the rest in this section. Recent measurements show that the upper critical field \cite{PhysRevLett.106.207001} and the electronic specific heat \cite{PhysRevLett.112.067002} of CeCu$_2$Si$_2$ under pressure follow two distinct scaling behaviors, supporting the scenario of  different order parameters in the two superconducting regimes.

\begin{figure}[t]
\includegraphics[angle=0,width=0.48\textwidth]{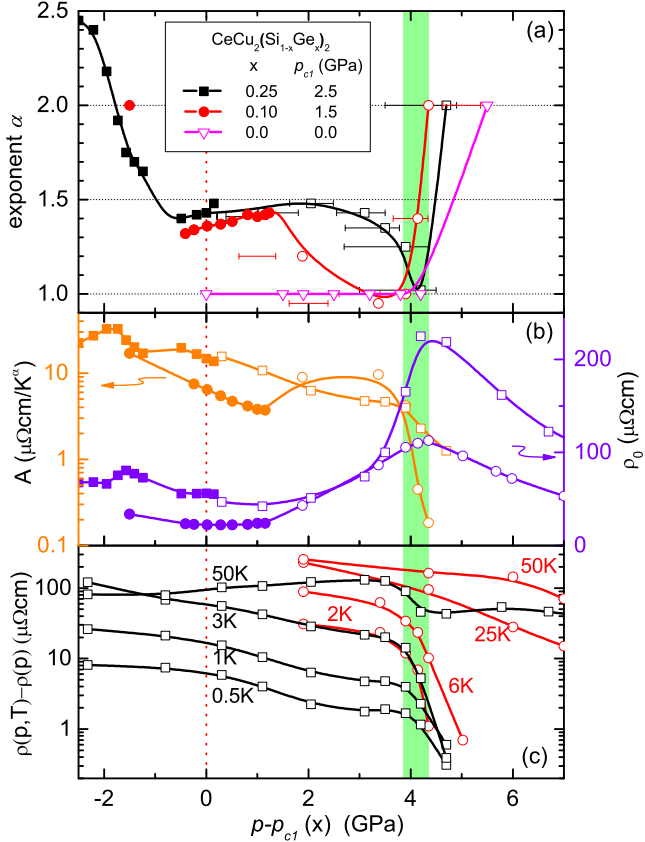}

\caption{ Pressure dependence of (a) resistivity exponent $\alpha$, (b) $A$ coefficient and residual resistivity $\rho_0$ and (c) resistivity isotherms $\Delta\rho_T(p)$ for CeCu$_2$(Si$_{1-x}$Ge$_x$)$_2$ with $x~=~0, 0.1$ and 0.25.  Adapted with permission from Ref.~\cite{YuanPRL}. Copyright 2006 the American Physical Society.}\label{F4yuanPRL}
\end{figure}

In the high-pressure region, the superconducting temperature $T_c$ of CeCu$_2$(Si$_{1-x}$Ge$_x$)$_2$ reaches a maximum value near a second critical point $p_{c2}$ which is associated with a possible first-order valence transition of the Ce ions \cite{Yuan19122003,YuanPRL}. As shown in Fig.~\ref{F4yuanPRL} \cite{YuanPRL}, a few pronounced features are observed near $p_{c2}$, (1) a linear temperature dependence of the electrical resistivity above $T_c$, (2) a drop of the electronic scattering as reflected in the resistivity coefficient $A$ and (3) a pronounced maximum of the residual resistivity. Furthermore, a maximum is also observed near $p_{c2}$ in the Sommerfeld coefficient $\gamma(p)$ \cite{Holmes04PRB}. All these experimental facts are consistent with theoretical predictions based on a generalized periodic Anderson model as described in Section II and support the proposal that superconductivity in the high-pressure regime is attributed to valence fluctuations. However, there is no direct experimental evidence showing a pressure-induced valence transition in CeCu$_2$Si$_2$ and CeCu$_2$Ge$_2$. For CeCu$_2$Ge$_2$, measurements of angle dispersive X-ray diffraction reveal a weak drop of the unit cell volume at a pressure of $p_{c2}\simeq 15$~GPa at $10$~K, which suggests a transition in the Ce valence \cite{Onodera2002113}. However, this anomaly needs be confirmed by further experiments. Recent x-ray absorption spectroscopy measurements demonstrate that the Ce valence in CeCu$_2$Si$_2$  changes continuously with increasing pressure at low temperatures \cite{PhysRevLett.106.186405, PhysRevLett.113.086403}. This suggests that if there is a first-order valence transition, the end point must lie at low temperatures. Furthermore, in CeCu$_2$(Si$_{1-x}$Ge$_x$)$_2$, a sharp drop is observed in the isothermal pressure dependence of the electrical resistivity near $p_{c2}$, which becomes broad with increasing temperature and vanishes at around $10$~K for $x=0.1$ and around $20$~K for $x=0.25$ \cite{YuanPRL}. A more recent analysis of the isothermal resistivity of CeCu$_2$Si$_2$ suggests that the valence transition reaches a critical end point around $p_{c2}=4.5$~GPa, at a negative temperature \cite{PhysRevLett.109.046401}. These transport measurements indicate that there may be a valence transition at very low temperatures but to date it has been challenging to detect this in x-ray based experiments.

\begin{figure}[t]
\includegraphics[angle=0,width=0.48\textwidth]{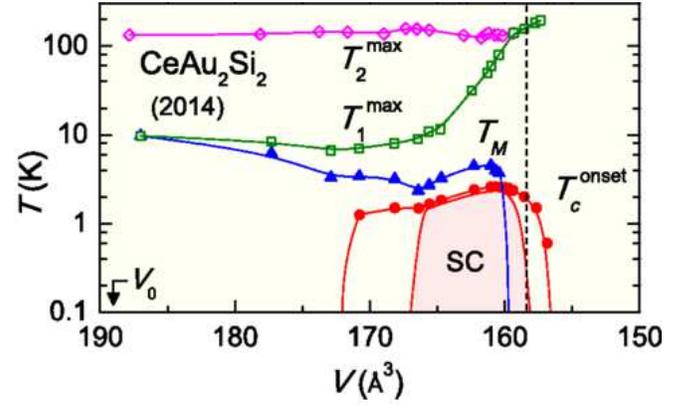}

\caption{ Temperature-volume phase diagram of CeAu$_2$Si$_2$, where $T_1^{max}$ and $T_2^{max}$ correspond to maxima in the resistivity, and $T_M$ and $T_c$ correspond to the magnetic ordering and superconducting transition temperatures respectively. Adapted with permission from Ref.~\cite{PhysRevX.4.031055}. Copyright 2014 the American Physical Society.}\label{F3cRen}
\end{figure}

Recently, pressure-induced superconductivity was also found in  antiferromagnetic CeAu$_2$Si$_2$, another compound isostructural and isoelectronic to CeCu$_2$Si$_2$ \cite{PhysRevX.4.031055}. Unlike CeCu$_2$Si$_2$, the phase diagram in Fig.~\ref{F3cRen} shows that superconductivity and antiferromagnetism coexist over a broad pressure range in CeAu$_2$Si$_2$, showing a parallel increase of $T_c$ and $T_M$ in the pressure range of 16.7~GPa $<p<$ 22.3~GPa, where $T_M$ possibly corresponds to a magnetic transition. This behavior again challenges the widely accepted proposal of spin fluctuation mediated superconductivity in heavy fermion compounds and the relationship between superconductivity and magnetism in the high pressure region remains unclear. It is important to find out whether the anomaly observed above $T_c$ is a continuous evolution of the antiferromagnetic transition observed at low pressures. We note that $T_N$ of CeAu$_2$Si$_2$ shows a minimum around $p\simeq16$~GPa, at which one can not exclude the possible suppression of the ambient-pressure antiferromagnetic transition.

It has also been proposed that fluctuations associated with  orbital transitions can arise in heavy fermion systems \cite{HattoriOrbFluc}. Using dynamical mean-field theory (DMFT), it was predicted that in CeCu$_2$Si$_2$ under pressure, there is a transition from the Ce $4f$ electron predominantly occupying the atomic CEF ground state, to it mainly occupying one of the excited states \cite{CCSOrbPred}. It  was further suggested that the fluctuations corresponding to this transition may lead to the superconducting dome at higher pressures. Orbital transitions were also predicted to occur under pressure in both CeAu$_2$Si$_2$ and  CeCu$_2$Ge$_2$ \cite{PhysRevX.4.031055},  and it was suggested that the different temperature-pressure phase diagram of  CeAu$_2$Si$_2$ arises due to different occupations of the CEF levels by the $4f$ electron. However, evidence for this scenario has not yet been found experimentally and there appear to be discrepancies between measurements such as nonresonant inelastic x-ray scattering (NIXS) and the predictions from DMFT \cite{CCGOrbAbs}. As a result, further experimental and theoretical efforts are necessary in order to elucidate the role of various degrees of freedom in the superconductivity of these compounds.

\begin{figure}[t]
\includegraphics[angle=0,width=0.48\textwidth]{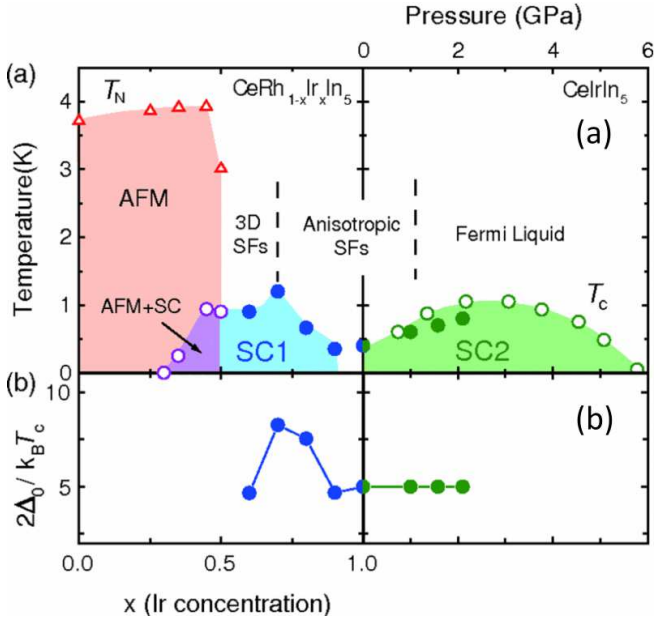}

\caption{ (a) Combined phase diagram of both CeRh$_{1-x}$Ir$_x$In$_5$ and CeIrIn$_5$ as a function of Ir concentration $x$ and pressure, showing the presence of magnetism and two superconducting domes. (b) The magnitude of the superconducting gap $2\Delta/k_BT_c$ across the superconducting regions, as a function of $x$. Adapted with permission from Ref.~\cite{PhysRevLett.96.147001}. Copyright 2006 the American Physical Society.}\label{F5Kawasaki}
\end{figure}

Besides CeCu$_2$Si$_2$ and similar compounds, the presence of two-superconducting domes has also been suggested for Ce(Rh,Ir)In$_5$ \cite{PhysRevLett.96.147001, PhysRevB.70.020505} and the Pu$T$In$_5$ compounds \cite{0953-8984-24-5-052206}. In CeIrIn$_5$,  $^{115}$In-NQR experiments demonstrate that the antiferromagnetic fluctuations are strongly suppressed with increasing pressure, but $T_c$ increases, reaching a maximum value of about 1~K at around 3~GPa \cite{PhysRevLett.94.037007}. On the other hand, substituting  Ir with Rh gives rise to antiferromagnetic order and enhances $T_c$ \cite{PhysRevB.64.100503}, with coexistence of superconductivity and magnetism in CeRh$_{1-x}$Ir$_x$In$_5$ for 0.35 $<x<$ 0.5 \cite{PhysRevLett.96.147001}. These results suggest that much like CeCu$_2$(Si$_{1-x}$Ge$_x$)$_2$ under pressure \cite{Yuan19122003}, two distinct superconducting states may exist in Ce(Rh,Ir)In$_5$ in the lattice density-temperature phase diagram, as shown in Fig.~\ref{F5Kawasaki} \cite{PhysRevLett.96.147001}. Here CeIrIn$_5$ was assumed to be located in the second superconducting dome which had been taken to be close to a charge (or valence) instability \cite{PhysRevLett.96.147001,PhysRevLett.94.037007}. However, our recent studies of substituting Hg and Sn on the In sites, and Pt on the Ir sites indicate that CeIrIn$_5$ is very much close to an AFM QCP, suggesting that spin fluctuations play a critical role in the formation of superconductivity \cite{PhysRevB.89.041101}. Furthermore, although there is a change in the orbital anisotropy of the ground-state wave function at lower values of $x$ in CeRh$_{1-x}$Ir$_x$In$_5$, a significant difference is not seen between $x~=~0.75$ and $x~=~1$, suggesting the lack of an orbital transition when crossing between the two superconducting domes \cite{Willers24022015}.   In the Pu$T$In$_5$ compounds, it was proposed that PuCoIn$_5$ is in the antiferromagnetic regime, while PuRhIn$_5$ and PuCoGa$_5$ are located in the second superconducting dome \cite{0953-8984-24-5-052206}. Recently, measurements of resonant ultrasound spectroscopy provide evidence for a softening of the bulk modulus at temperatures above $T_c$ in PuCoGa$_5$, which is characteristic of valence instabilities \cite{0953-8984-24-5-052206}. This suggests that valence fluctuations may play a crucial role in the unusually high $T_c$ ($\simeq 18.5$~K) of PuCoGa$_5$. Since a similar enhancement of $T_c$ is also observed in CeCu$_2$(Si$_{1-x}$Ge$_x$)$_2$ under pressure, critical valence fluctuations may favor superconductivity at higher temperatures. However, further experiments are needed to confirm the existence of this pairing mechanism in these compounds.

A pressure-induced volume collapse transition, which is usually associated with the depopulation of localized $f$~electrons, has been observed in a number of rare earth elements and compounds. For examples, a pronounced volume collapse transition was shown to exist in the rare earth elements Ce, Pr and Gd under pressure and at room temperature \cite{McMahan1998}. Similar isostructural electronic transitions were also observed in several intermetallic compounds, such as rare earth monochalcogenides \cite{PhysRevB.6.2285} and YbInCu$_4$ \cite{de_teresa_pressure_1996}. In YbInCu$_4$, the valence transition is suppressed by applying pressure, and a magnetic-like transition develops in the critical pressure regime \cite{PhysRevB.58.409,PhysRevLett.96.046405}. Recently, evidence for a pronounced anomaly in the valence under pressure was also found in Yb$_2$Ni$_{12}$As$_7$ at around 1~GPa \cite{jiang_crossover_2015}. However, superconductivity has not yet been observed in most of these compounds showing a pronounced valence transition and the conditions which favor superconductivity mediated via the valence-fluctuation mechanism are unclear. In CeCu$_2$Si$_2$ and similar compounds, we note that there is a weak first-order valence transition with a low-lying critical end point \cite{YuanPRL, PhysRevLett.106.186405} and it may be that these conditions are necessary for such a superconducting state to form.

\begin{figure*}[t]
\begin{center}
\includegraphics[angle=0,width=0.5\textwidth]{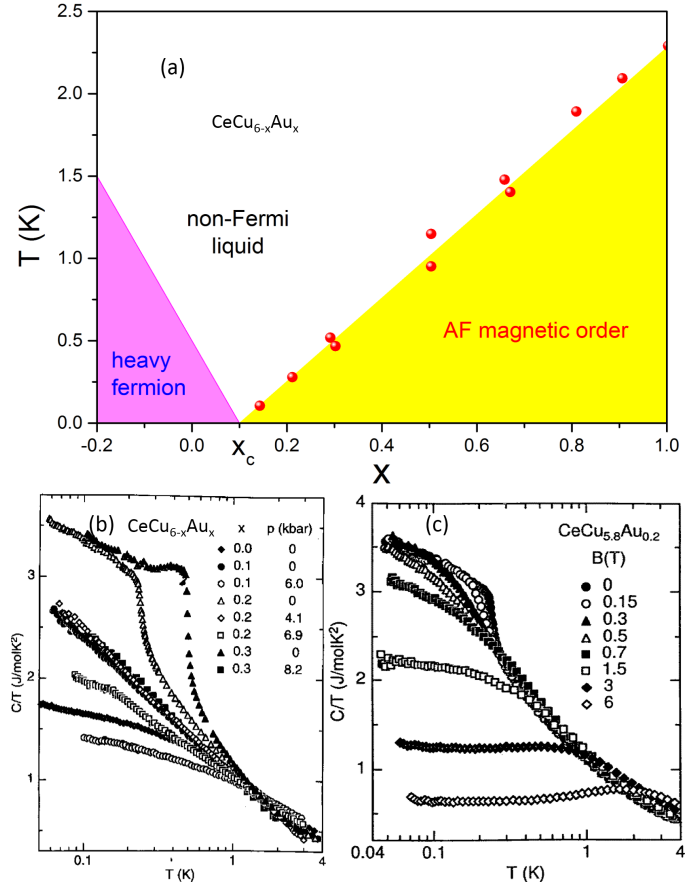}
\end{center}
\caption{  (a) $T-x$ phase diagram of ${\mathrm{CeCu}}_{6-x}{\mathrm{Au}}_{x}$ \cite{0953-8984-8-48-003}. The negative values of $x$ in the phase diagram correspond to applying a positive pressure. (b) Temperature dependence of the specific heat of ${\mathrm{CeCu}}_{6-x}{\mathrm{Au}}_{x}$ for different doping levels and pressures. (c) Temperature dependence of the specific heat of ${\mathrm{CeCu}}_{5.8}{\mathrm{Au}}_{0.2}$ in different applied magnetic fields. (b) and (c) are adapted with permission from Ref.~\cite{0953-8984-8-48-003}. Copyright 1996 the Institute of Physics.}\label{lohne}
\end{figure*}

Upon tuning a system from the heavy fermion to intermediate valence state by increasing the hybridization strength, various degrees of freedom may play important roles in the critical behavior and/or the development of superconductivity. Whether the spin, charge or orbital ordering of the $f$~electrons undergo a critical instability simultaneously or independently is currently unclear. The above discussion demonstrates that the spin and charge channels are sometimes well separated under pressure, showing multiple quantum phase transitions with distinct critical behaviors. Further experimental and theoretical studies are required to elucidate the superconducting pairing states and to understand the critical behavior associated with the instabilities of the various degrees of freedom in heavy fermion systems.

\section{Multiple quantum phase transitions}

The competing phases in heavy fermion compounds  typically have a low energy scale, which can be easily tuned by external parameters. Thus, heavy fermion compounds, in particular those based on the elements of Ce and Yb, have been regarded as prototype materials for studying quantum phase transitions. In the preceding sections, we have described superconductivity emerging near magnetic and charge instabilities, which are associated with various degrees of freedom of the $f$~electrons. In this section, we discuss multiple magnetic quantum phase transitions reached upon  tuning different parameters, such as pressure, magnetic field and chemical doping, and examine their universal classification.

Several theories of quantum criticality have been proposed for heavy fermions compounds \cite{PhysRevB.14.1165Hertz, PhysRevB.48.7183Millis, Moriya, si_locally_2001, ColemanPepin,Pepin2007,PhysRevB.84.041101,PhysRevLett.105.186403}. Even though the underlying physics at QCPs is still under debate, the following two scenarios have been widely discussed in the literature. The first scenario is the conventional SDW-type of QCP \cite{PhysRevB.14.1165Hertz, PhysRevB.48.7183Millis, Moriya, DPines07Nature}, at which the quasiparticles are scattered by critical SDW spin fluctuations, leading to an increased effective mass and a continuous evolution of the Fermi surface across the QCP. It appears that most AFM QCPs where superconductivity is observed are of the SDW-type. At an unconventional local QCP \cite{si_locally_2001, ColemanPepin}, critical fermionic and bosonic degrees of freedom are involved and the Fermi volume undergoes a sudden change as a result of the breakdown of the Kondo effect. As a result, the evolution of the Fermi surface may be used to characterize a QCP and therefore to test theoretical models.

Evidence for various types of critical behavior have been reported in heavy fermion compounds \cite{StewartNFL,RevModPhys.79.1015} but the universal classification could be complicated by extrinsic factors such as disorder, the measured temperature range and additional symmetry breaking from applied magnetic fields. Furthermore, different scaling behaviors are found in some compounds when different parameters are tuned. The classification of QCPs and the question of whether they show universal behavior remains an open issue. Here we will focus on a few examples where the ground state can be tuned by different parameters.

\begin{figure}[t]
\includegraphics[angle=0,width=0.48\textwidth]{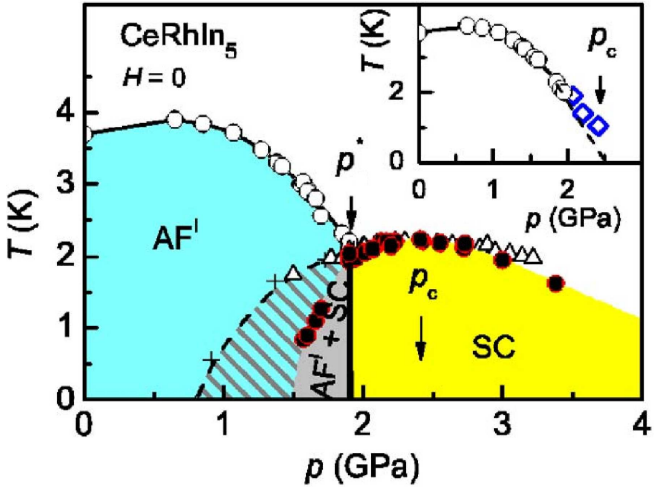}

\caption{ Temperature-pressure phase diagram of CeRhIn$_5$ Pressure suppresses the magnetism until $T_N$ is equal to $T_c$ at $p^*$. The inset shows the pressure dependence of $T_N$ in a large enough field, which shows $T_N$ suppressed to zero at $p_c$. Adapted with permission from Ref.~\cite{PhysRevB.74.020501}. Copyright 2006 the American Physical Society.}\label{F2Knebel}
\end{figure}

\begin{figure}[t]
\includegraphics[width=0.4\textwidth]{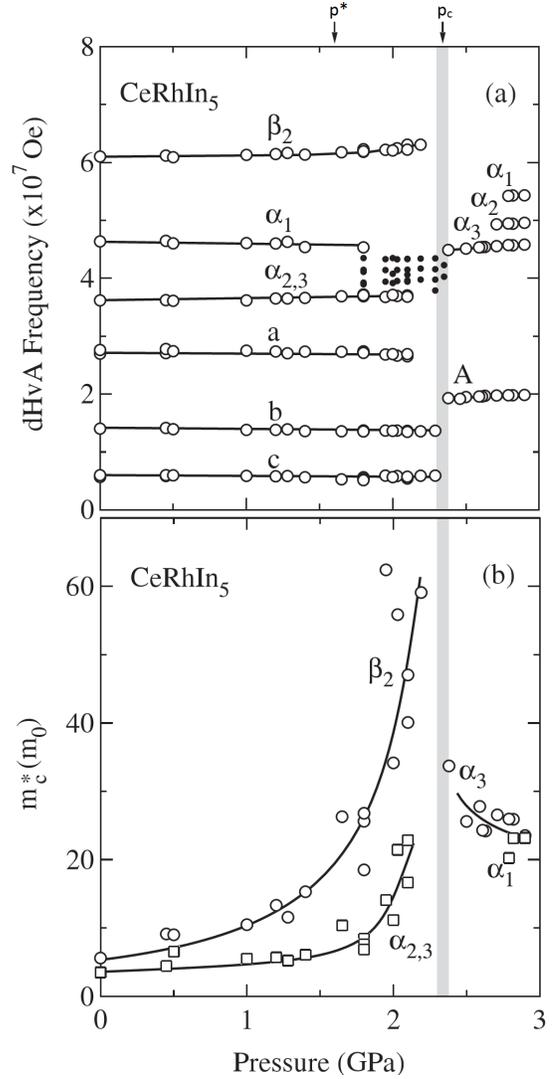}
\vspace{-10pt}
\caption{ Pressure dependence of dHvA measurements of CeRhIn$_5$, showing (a) dHvA frequencies and (b) the cyclotron masses. A sudden change of frequencies is seen at $p_c$, where there is also a sharp upturn of the masses. Adapted with permission from Ref.~\cite{10.1143/JPSJ.74.1103}. Copyright 2005  the Physical Society of Japan}\label{F4Shishido}
\end{figure}

${\mathrm{CeCu}}_{6}$ is an archetypal heavy fermion compound with a large specific heat coefficient $\gamma$ = 1.6~J/(mol~K$^2$) \cite{doi:10.1143/JPSJ.54.304} and a Kondo temperature $T_K$ of around 6~K \cite{:/content/aip/journal/jap/57/8/10.1063/1.335212}. No magnetic order is observed in ${\mathrm{CeCu}}_{6}$ down to the lowest measured temperatures and the electrical resistivity follows a quadratic temperature dependence, which is a characteristic feature of Fermi liquid behavior. Substituting Au for Cu results in a lattice expansion which reduces $J$, and as a result  long range antiferromagnetic order emerges in ${\mathrm{CeCu}}_{6-x}{\mathrm{Au}}_{x}$ for $x > x_c \simeq 0.1$ \cite{stockert_unconventional_2011}, as shown in Fig.~\ref{lohne}(a). Around the AFM QCP ($x_c=0.1$), the system shows pronounced non-Fermi liquid behavior. For instance, the electrical resistivity follows a linear temperature dependence and the specific heat $C/T$ diverges logarithmically with decreasing temperature \cite{0953-8984-8-48-003}, as shown in Fig.~\ref{lohne}(b). The Kondo temperature, as investigated by ultraviolet and x-ray photoemission spectroscopy, shows a step-like change at $x_c$ \cite{PhysRevLett.101.266404, PhysRevB.79.075111, 0953-8984-22-16-164203}, possibly indicating the destruction of heavy quasiparticles. Moreover, neutron scattering experiments also revealed anomalous $E/T$ and $H/T$ scaling of the susceptibility \cite{schroder_onset_2000}, suggesting the development of an unusual kind of localized excitation. These experimental facts support the presence of a unconventional QCP at the critical doping concentration of $x_c$. Similar critical behavior is also observed when the AFM order of CeCu$_{5.8}$Au$_{0.2}$ ($T_N\simeq 250$~mK) is suppressed by pressure at $p_c~=~4$~GPa \cite{PhysRevB.63.134411}. Meanwhile, the application of a small magnetic field ($B_c \simeq 0.4$~T for $H//c$) suppresses the AFM order in CeCu$_{5.8}$Au$_{0.2}$ and the critical behavior at the field-tuned QCP differs from that induced by pressure or doping  \cite{PhysRevB.63.134411}. Inelastic neutron scattering measurements demonstrate that the critical fluctuation spectrum of CeCu$_{5.8}$Au$_{0.2}$ near the critical field $B_c$ is better characterized by the three dimensional SDW scenario rather than the unconventional local  scenario found in CeCu$_{5.9}$Au$_{0.1}$ at zero magnetic field \cite{PhysRevLett.99.237203}. Moreover, at the field-induced QCP the specific heat and electrical resistivity follow $C/T \propto -T^{0.5}$ (Fig.~\ref{lohne}(c)) and $\rho\propto T^{3/2}$ dependencies respectively, which suggest three dimensional SDW-type quantum criticality. These results indicate that different types of critical behavior can be induced by pressure and magnetic field and multiple QCPs may exist in the same compound. Besides CeCu$_{6-x}$Au$_x$, evidence for QCPs with different behaviors exist for other compounds. For example, there is a growing body of experimental evidence suggesting an unconventional local QCP in YbRh$_2$Si$_2$ under applied magnetic field \cite{YRSQCP1,YRSQCP2,paschen_hall-effect_2004}, although alternative theoretical models have also been proposed \cite{PhysRevB.84.041101, Miyake, Vojta}. On the other hand, chemical pressure via the substitution of Ir or Co leads to the separation of the AFM QCP and Kondo breakdown line \cite{YRSNat2009}, and in the case of Ir doping, it was suggested that a spin-liquid-type ground state may emerge. However, we will mainly focus on Ce-based heavy fermion compounds and an extensive discussion of other compounds is beyond the scope of this article.

\begin{figure}[t]
\includegraphics[angle=0,width=0.48\textwidth]{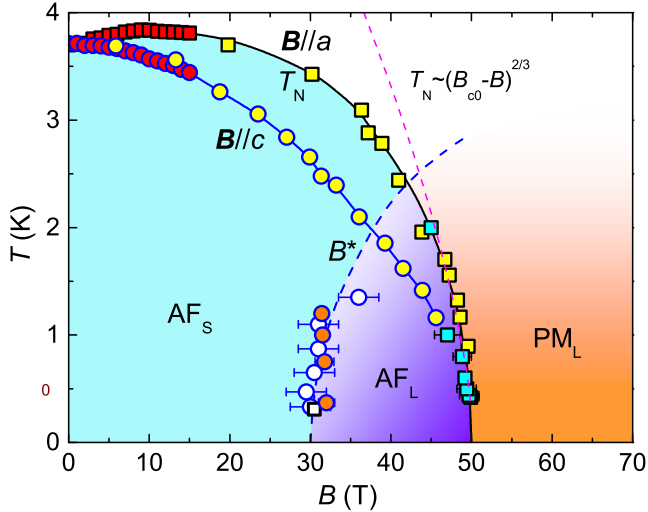}

\caption{ Temperature - magnetic field phase diagram of CeRhIn$_5$ at ambient pressure. The line $B^*$ separates the antiferromagnetic phases with large (AF$_L$) and small (AF$_S$) Fermi surface volumes.  Adapted with permission from Ref.~\cite{Jiao20012015}. Copyright 2015 the National Academy of Sciences of the United States of America.}\label{F5Ajiao}
\end{figure}

In order to search further evidence for multiple QCPs and to study their classification, we have recently extended measurements of CeRhIn$_5$  to very high magnetic fields and pressure, which have revealed remarkable new physics. CeRhIn$_5$ belongs to the well-investigated Ce$T$In$_5$ ($T$ = Co, Rh, Ir) family of compounds. At ambient pressure, CeRhIn$_5$ orders antiferromagnetically below $T_N=3.8$~K \cite{PhysRevLett.84.4986}, but CeCoIn$_5$ and CeIrIn$_5$ are heavy fermion superconductors \cite{CPetrovic01CeCoIn5, 0295-5075-53-3-354}. Previous dHvA measurements on La-substituted Ce$T$In$_5$ samples have shown that the Ce-$4f$ electrons are itinerant in CeCoIn$_5$ and CeIrIn$_5$ but localized in CeRhIn$_5$ \cite{doi:10.1143/JPSJ.71.162, PhysRevLett.93.186405}. This conclusion is further supported by a comparison of the experimentally derived dHvA frequencies with band structure calculations under the assumption that the $4f$ electrons are either localized or itinerant\cite{doi:10.1143/JPSJ.72.854, Jiao20012015}. These results demonstrate that in ambient conditions CeRhIn$_5$ possesses a small Fermi surface volume, but in CeCoIn$_5$ the $f$~electrons contribute to the Fermi sea and therefore the volume of the Fermi surface is enlarged. Such an expansion of the Fermi volume may correspond to the delocalization of Ce-$4f$ electrons with increasing the hybridization strength.

The application of hydrostatic pressure suppresses the antiferromagnetic order in CeRhIn$_5$ and induces superconductivity \cite{park_hidden_2006, PhysRevB.74.020501}, as shown in Fig.~\ref{F2Knebel}. Superconductivity and magnetism coexist below $p^*\simeq 1.75$~GPa. In the pressure range  $p^*<p<p_c=2.3$~GPa, magnetism is hidden inside the superconducting state at $B=0$ and long-range magnetic order is recovered by applying a magnetic field. This allows the tracking of the pressure-induced magnetic QCP at $p_{c}\simeq 2.3$~GPa\cite{park_hidden_2006}. As shown in Fig.~\ref{F4Shishido}, a pressure-induced dramatic change of the dHvA frequencies and a divergence of the effective mass are observed in CeRhIn$_5$ at the AFM QCP \cite {doi:10.1143/JPSJ.71.162}, which have been regarded as experimental evidence to support the unconventional local QCP scenario \cite{Si1161}. On the other hand, the antiferromagnetic order is robust against magnetic field \cite{doi:10.1143/JPSJ.71.162}, allowing us to study the evolution of the Fermi surface with applied field. This is not possible for CeCu$_{6-x}$Au$_x$ and YbRh$_2$Si$_2$ due to the  small values of the critical fields. Therefore, CeRhIn$_5$ may be a model system to study  multiple QCPs and their relationship to the  Fermi surface reconstruction, which motivated us to systematically study the properties of CeRhIn$_5$ under the extreme conditions of high magnetic field and high pressure, using high-field facilities.

\begin{figure}[t]
\includegraphics[angle=0,width=0.48\textwidth]{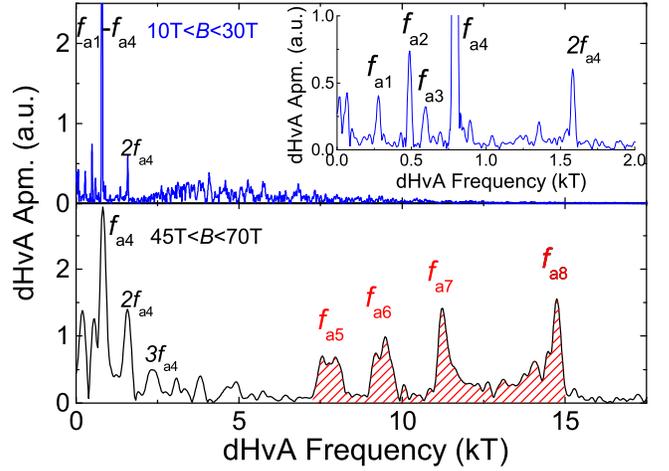}

\caption{ Fourier spectra of the dHvA oscillations of CeRhIn$_5$ at ambient pressure, over field windows of 10~T $< B <$ 30~T (top) and 45~T $< B <$ 70~T (bottom). In the high field region, four new dHvA frequencies are observed.  Adapted with permission from Ref.~\cite{Jiao20012015}. Copyright 2015 the National Academy of Sciences of the United States of America.}\label{F4Bjiao}
\end{figure}

Measurements of the ac specific heat and magnetic susceptibility allowed us to map the $B-T$ phase diagram of CeRhIn$_5$ (Fig.~\ref{F5Ajiao}) \cite{Jiao20012015}. It can be seen that $T_N$ is eventually suppressed with increasing magnetic field for $H//a$ and $H//c$, which converge to the same critical field at the AFM QCP ($B_{c0}\simeq 50$~T). It is remarkable that the magnetic order is suppressed at nearly the same critical field for the two different field orientations, since this compound possesses a quasi-two dimensional crystal structure. Furthermore, near $B_{c0}$ the behavior follows $T_N\thicksim(B_{c0}-B)^{2/3}$ ($H//a$), which indicates three dimensional spin fluctuations.

\begin{figure}[t]
\includegraphics[angle=0,width=0.48\textwidth]{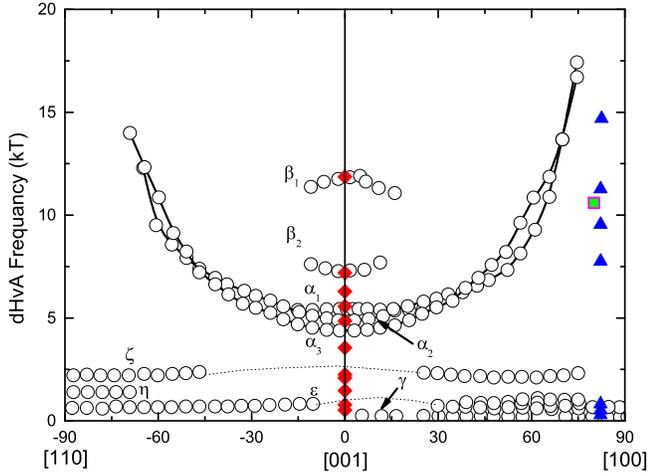}

\caption{  Comparison of the dHvA frequencies of CeRhIn$_5$ at $B > B^*$ and CeCoIn$_5$ (black open circles). The similarity between the two materials suggests that the $4f$ electrons of CeRhIn$_5$  are delocalized at high magnetic fields. Adapted with permission from Ref.~\cite{Jiao20012015}. Copyright 2015 the National Academy of Sciences of the United States of America.}\label{FS6Jiao}
\end{figure}

More interestingly, a field-induced reconstruction of the Fermi surface is observed around $B^*\simeq 30$~T and at temperatures below 2~K \cite{Jiao20012015}, which is well inside the AFM state of CeRhIn$_5$ (see Fig.~\ref{F5Ajiao}). Evidence for a pronounced change of the dHvA frequencies was first found in  dHvA measurements up to applied fields of 72~T, using a pulsed magnetic field at Los Alamos National Laboratory, as shown in Fig.~\ref{F4Bjiao} \cite{Jiao20012015}. This important observation was later confirmed by measurements of the torque magnetization and Hall resistivity \cite{Jiao20012015}, which clearly show a change around $B^*\simeq 30$~T for both $H//c$ and $H//a$. For $B>B^*$, new dHvA frequencies with much larger values are observed, suggesting an expansion of the Fermi volume above $B^*$. Moreover, we found that the dHvA frequencies of CeRhIn$_5$ in the field range $B>B^*$ are very much compatible with those of CeCoIn$_5$ (see Fig.~\ref{FS6Jiao}), indicating that the Ce-$4f$ electrons become delocalized at $B>B^*$ \cite{Jiao20012015}. By fabricating a microstructure of CeRhIn$_5$ using the focused ion beam technique, Moll $\textit{et al}$ found  hysteresis around $B^*\simeq 30$~T in the in-plane resistivity with the field applied close to the $c$-axis, where a Fermi surface reconstruction is also observed \cite {moll_field-induced_2015}. They argued that a density wave transition may occur at $B^*$ \cite {moll_field-induced_2015}. However, our recent measurements of the magnetostriction down to 0.6~K using a pulsed magnetic field do not show any feature at $B^*~=~30$~T \cite{unpublished}, indicating that the transition at $B^*$ is unlikely to be driven by electron-phonon coupling, as would be the case for charge density wave order.

\begin{figure}[t]
\includegraphics[angle=0,width=0.48\textwidth]{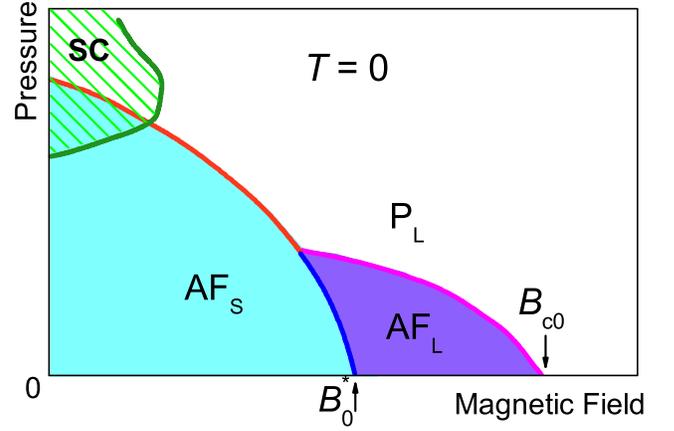}

\caption{ Schematic pressure-field phase diagram of CeRhIn$_5$ at zero temperature, where different behaviors of the QCPs are seen when pressure and magnetic fields are applied. Adapted with permission from Ref.~\cite{Jiao20012015}. Copyright 2015 the National Academy of Sciences of the United States of America.}\label{F5BJiao}
\end{figure}

Fig.~\ref{F5BJiao} shows a schematic $p-B$ phase diagram of CeRhIn$_5$ at zero temperature \cite{Jiao20012015}. The pressure-induced AFM QCP is characterized by a dramatic change of the Fermi surface from the AFM state with a small volume (AF$_S$) to the paramagnetic Fermi surface with a large volume (P$_L$) upon increasing the pressure. These observations seem to support the scenario of a local QCP \cite{si_locally_2001,Si1161}, with the occurrence of superconductivity near the QCP. On the other hand, upon suppressing the AFM state by increasing the magnetic field, a reconstruction of the Fermi surface occurs  at $B^*$, deep inside the AFM state of CeRhIn$_5$ \cite{Jiao20012015}. The measurements of quantum oscillations and Hall resistivity, together with band structure calculations, indicate there is a possible Kondo breakdown transition at $B^*$, which leads to a change from a small Fermi surface volume (AF$_S$) to a large Fermi surface volume (AF$_L$)  \cite{Jiao20012015}. At the magnetic field-induced QCP, the Fermi surface evolves smoothly from the AFM state to the paramagnetic state, indicating it is of the SDW-type. This is also consistent with the  critical point being the same for fields along the $a$- and $c$-axes. From this phase diagram, one can see that multiple quantum phase transitions can be induced by tuning different parameters in the same compound. Furthermore, two distinct QCPs may exist under magnetic field in CeRhIn$_5$, a Kondo breakdown QCP at $B^*$ and a SDW QCP at $B_{c0}$. These experimental findings not only provide new perspectives on CeRhIn$_5$, but also suggest that the change of the Fermi surface volume may allow the classification of the different  scenarios proposed for QCPs. To test the universality of this behavior, it is important to map the $B-p$ phase diagram of CeRhIn$_5$ and extend these investigations to other heavy fermion systems, with localized $f$ electrons.

\begin{figure}[t]
\includegraphics[angle=0,width=0.48\textwidth]{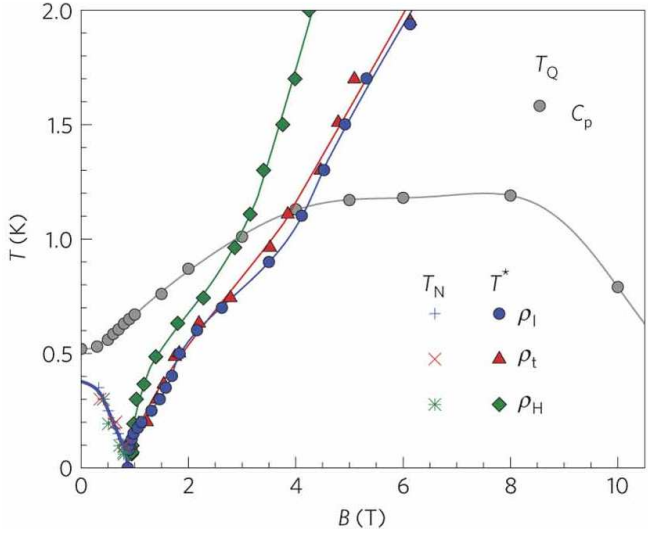}

\caption{ Temperature-field phase diagram of ${\mathrm{Ce}_3\mathrm{Pd}_{20}\mathrm{Si}_6}$, which shows the antiferromagnetic ordering temperature $T_N$, the quadrupolar ordering temperature $T_Q$ and $T^*$, which characterizes the crossover from localized to itinerant behavior of the $4f$ electrons. Adapted with permission from Ref.~\cite{custers_destruction_2012}. Copyright 2012 the Nature Publishing Group.}\label{F3aCusters}
\end{figure}

Very recently, differences were also observed between the pressure- and field- tuning of ${\mathrm{CeNi}}_{2-\sigma}{\mathrm{As}}_{2}$($\sigma \simeq$ 0.28) \cite{Luo03112015}, which crystallizes in a ${\mathrm{CaBe}}_{2}{\mathrm{Ge}}_{2}$-type structure with $T_N\simeq 5.1$~K. This compound shows semimetallicity with a low charge carrier density and a moderate degree of magnetic frustration. The application of pressure suppresses the AFM transition, with a critical pressure near $p_c=2.7$~ GPa derived from ac specific heat measurements. However, an upturn is observed in the electrical resistivity at temperatures below 2~K in the pressure range 2.02~GPa $<p<3.5$~GPa, the origin of which is unknown. At low pressures ($p<p_c$), a step like jump is observed in the Hall resistivity $\rho_{xy}$, which is explained by a spin-flop transition. The coherence temperature seems to vanish at $p_c$, for which a scenario of a Kondo destruction QCP was proposed. However, there is still little evidence for either a Fermi surface change or a jump in the charge carrier density at $p_c$. On the other hand, the antiferromagnetic order is also suppressed by applying a magnetic field of $B_c=2.8$~T, where the field-induced QCP is assumed to be close to the quantum critical end point of a spin-flop transition.

\begin{figure}[t]
\includegraphics[angle=0,width=0.49\textwidth]{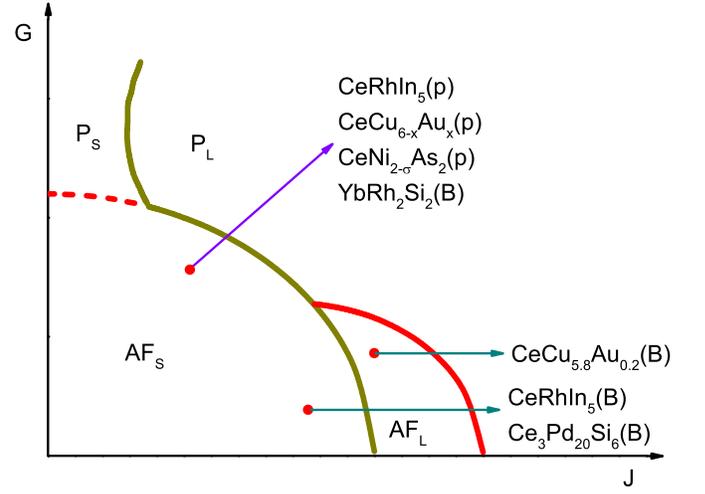}

\caption{\label{FigGlobalphasediagram} Global phase diagram for heavy fermion systems \cite{Si2006Global,Coleman2010Global}.  The parameter $J$ is the coupling strength, while $G$ characterizes the degree of frustration. The phases AF$_S$ and AF$_L$ are magnetically ordered with large and small Fermi surface volumes respectively, while P$_S$ and P$_L$ are the respective paramagnetic phases. Schematic trajectories for different materials are displayed based on experimental results obtained via tuning pressure ($p$) or applied magnetic fields (B).}\label{Globalphasediagram}
\end{figure}

The caged compound ${\mathrm{Ce}_3\mathrm{Pd}_{20}\mathrm{Si}_6}$ is another example useful for studying the effects of the dimensionality on  quantum criticality \cite{custers_destruction_2012}. This compound crystallizes in a cubic structure with space group $Fm\bar{3}m$ and carries a huge specific heat coefficient of $\gamma$ = 8~J/(mol~K$^2$). As shown in Fig.~\ref{F3aCusters}, with decreasing temperature, ${\mathrm{Ce}_3\mathrm{Pd}_{20}\mathrm{Si}_6}$ undergoes two subsequent phase transitions at 0.53~K and 0.33~K, which correspond to quadrupolar and antiferromagnetic transitions respectively. The quadrupolar transition temperature $T_Q$ first increases with the application of magnetic fields and peaks at 8~T, with $T_Q\simeq$ 1.2~K, before decreasing at larger fields. Further increasing the magnetic field will suppress $T_Q$ completely, implying a possible field-induced quantum critical point above 10~T. Inside the quadrupolar phase, the antiferromagnetic transition temperature $T_N$ is suppressed monotonically by applying a field and reaches zero at a critical field around 0.9~T. In an applied field of 1~T, near the critical field, the electrical resistivity shows a linear temperature dependence and the specific heat $C/T$ diverges logarithmically, but upon moving away from the critical field, Fermi liquid behavior is recovered. Moreover, the crossover temperature line $T^*(B)$, where there is a change of the Hall coefficient, merges with the $T_N(B)$ line at this critical field inside the quadrupolar ordered state, similar to the field-induced Kondo breakdown observed in CeRhIn$_5$ \cite{Jiao20012015}. It was suggested that the increased dimensionality of the three dimensional cubic crystal structure of ${\mathrm{Ce}_3\mathrm{Pd}_{20}\mathrm{Si}_6}$ compared to YbRh$_2$Si$_2$, means that the magnetic frustration in the global phase diagram is reduced and therefore the two compounds follow different paths.

To summarize this section, we plot a materials-based global phase diagram in Fig.~\ref{Globalphasediagram}. Pressure, magnetic field, doping and lattice dimensionality are possible tuning parameters for quantum criticality in heavy fermion compounds. From Fig.~\ref{Globalphasediagram}, three typical tuning routes from the AF$_S$  phase to P$_L$ can be considered \cite{Si2006Global,SiPSSGlobal}. Firstly the system can cross directly to the P$_L$ state, which corresponds to a local QCP. Secondly there may be a change of Fermi surface volume in the ordered state as the system goes from AF$_S$ to AF$_L$, before passing through a SDW QCP. Finally, the system may bridge the AF$_S$ and P$_L$ states by passing through the disordered P$_S$ phase. It can be seen that different scenarios for QCPs predicted in the global phase diagram can be realized either in different materials or in the same material with different tuning parameters. To classify the multiple quantum  phase transitions, the evolution of the Fermi surface is an important indicator. Besides the conventional SDW-type QCP and the unconventional local QCP which have been discussed frequently in the literature, the Kondo breakdown transition inside the magnetically ordered state may present another new scenario for quantum criticality, which should be explored. Furthermore, more experimental evidence is needed to elucidate the Kondo breakdown transition, and new materials will help establish the applicability of the global phase diagram.

\section{Summary and prospects}

In heavy fermions compounds, energy scales associated with the various degrees of freedom are well separated and the effects of strong electronic correlations can be studied without being complicated by other excitations, $\textit{e.g.}$, lattice dynamics. Moreover, the availability of very clean heavy fermion samples together with clean tuning parameters such as pressure allows us to study the fundamental physics without the influence of disorder. Applications of pressure, magnetic field and doping may enhance the hybridization between $f$ and conduction electrons, which can lead to a delocalization of the $f$~electrons, changing the ground state from antiferromagnetism to a heavy fermion paramagnet. Upon crossing the AFM QCP, unconventional normal and superconducting states emerge. When the hybridization is further increased, the $f$~electrons can hop between different valence states, which may also lead to novel behavior in the vicinity of a valence instability. Therefore, Ce-based heavy fermion compounds are prototype systems for studying unconventional superconductivity and multiple quantum phase transitions, which are induced either by tuning the magnetic state with different parameters or by tuning one parameter through instabilities of different degrees of freedom.

Intensive efforts have recently been devoted to the study of quantum phase transitions. However, the fate of quasiparticles at the QCP and the interplay between superconductivity and quantum criticality remains an open question. The  universal classification of QCPs is essential for establishing whether there is a global phase diagram for heavy fermions systems and the discovery of new  materials or further characterization of known compounds is necessary to test its applicability. Our recent measurements of the dHvA effect and Hall resistivity of CeRhIn$_5$ suggest that the evolution of the Fermi surface may provide a means for classifying QCPs, where the  destruction of the Kondo effect leads to a Fermi surface reconstruction. Nevertheless, besides the Kondo destruction scenario \cite{si_locally_2001, ColemanPepin}, alternative models have been proposed to understand the Fermi surface change at the AFM QCP \cite{PhysRevB.84.041101, Vojta, Miyake, Zhang2015}.

The identification of the symmetry of the order parameter is important for understanding the pairing mechanism of a superconductor. However, Ce based heavy fermion superconductors usually have quite a low superconducting transition temperature, typically of the order of $1$~K or lower, and in many cases superconductivity requires the application of pressure. These conditions largely restrict the range of experimental techniques which can be applied to characterize the superconducting order parameter. Except for Ce$T$In$_5$ compounds, the order parameters for most of other heavy fermion superconductors have not been thoroughly investigated and in several cases remain controversial. On the other hand, improvements of the experimental resolution of modern techniques and the availability of low-temperature conditions for a wider range of spectroscopic methods, such as ARPES and STM, may provide powerful tools to directly probe the momentum-dependence of the gap structure below $T_c$. Furthermore, spectroscopic measurements will also help reveal the development of Kondo screening as a function of temperature or other parameters in the normal state.

Heavy fermion superconductors where the crystal structure lacks inversion symmetry have also attracted considerable interest. The absence of inversion symmetry leads to an antisymmetric spin-orbit coupling (ASOC), allowing the admixture of spin-singlet and spin-triplet components in the superconducting pairing state \cite{NCSGorkov}. This mixed pairing state is prohibited in centrosymmetric superconductors, where the parity of the pairing state is a good quantum number. Evidence for nodal superconductivity but with a coherence peak in the spin lattice relaxation rate was observed in noncentrosymmetric CePt$_3$Si \cite{CePt3Si2004,CePt3SiNMR2}. For pressure-induced superconductivity in CeRhSi$_3$ \cite{CeRhSi3SC}, a huge upper critical field of around $B_{c2}\sim 30$~T was found for fields along the $c$ axis, with just 7~T in the $ab$~plane \cite{CeRhSi3Hc2}. These unusual superconducting properties might be related to the absence of inversion symmetry, but separating the effects of the ASOC from those of magnetism and strong electronic correlations are challenging. Superlattices consisting of layers of superconducting CeCoIn$_5$ separated by spacer layers of YbCoIn$_5$ \cite{2DHFSCRep}, where inversion symmetry is broken at the interface of the two materials provides a system where the ASOC can be tuned more controllably and indeed, different behavior of $H_{c2}(T)$ is found when the layer thickness is tuned \cite{SuperLattHc2a,SuperLattHc2b}. It may be the case that weakly correlated noncentrosymmetric superconductors are better suited for determining the effects of inversion symmetry breaking, for instance Li$_2$Pd$_3$B and Li$_2$Pt$_3$B \cite{Li2Pt3BNode}. A comprehensive review article on superconductivity  and spin-orbit coupling in non-centrosymmetric compounds will be available \cite{NCSRev}.

We would like to thank S. Arsenijevi$\acute{\mathrm{c}}$, E. D. Bauer, Y. Chen, D. Graf, N. Hussey, M. Jaime, S. Kirchner, Y. Kohama, H. O. Lee, K. Miyake, T. Park, Q. Si, J. Singleton, F. Steglich, J. D. Thompson, Z. Wang, J. Wosnitza, Y. J. Zhang and J.-X Zhu for useful discussions. This work was supported by Natural Science Foundation of China (grant numbers 11474251, 11374257 and 11174245),  the Science Challenge Program of China  and the Fundamental Research Funds for the Central Universities.

\bibliographystyle {apsrev}
\bibliography {RoPPRef-XLuV12}

\end{document}